\documentclass[aps,prb,superscriptaddress,twocolumn,showpacs]{revtex4-1}

\usepackage{amsmath, amsthm, amssymb, mathtools}
\usepackage{graphicx}
\usepackage[usenames,dvipsnames]{color}
\usepackage{times}
\usepackage{dsfont}
\usepackage{epstopdf}

\usepackage[sort&compress]{natbib}
\setcitestyle{numbers,square}
\usepackage{hyperref}
\hypersetup{
        colorlinks=true,
}

\usepackage[lofdepth,lotdepth,caption=false]{subfig}

\newcommand{\id}{\mathds{1}}

\newcommand{\be}{\begin{equation}}
\newcommand{\ee}{\end{equation}}
\newcommand{\bea}{\begin{eqnarray}}
\newcommand{\eea}{\end{eqnarray}}

\newcommand{\eqnref}[1]{(\ref{#1})}

\renewcommand{\i}{\mathfrak{i}}

\renewcommand{\vec}[1]{\mathbf{#1}}

\newcommand{\ba}{{\bf a}}

\newcommand{\bq}{{\bf q}}
\newcommand{\br}{{\bf r}}

\newcommand{\Ss}{\mathcal{S}}

\begin{document}

\title{Symmetry-protected non-Fermi liquids, Kagome spin liquids, \\ and the chiral Kondo lattice model}

\author{Bela Bauer}
\affiliation{Station Q, Microsoft Research, Santa Barbara, CA 93106-6105, USA}

\author{Brendan P. Keller}
\affiliation{Department of Physics, University of California, Santa Barbara, CA 93106, USA}

\author{Simon Trebst}
\affiliation{Institute for Theoretical Physics, University of Cologne, 50937 Cologne, Germany}

\author{Andreas W. W. Ludwig}
\affiliation{Department of Physics, University of California, Santa Barbara, CA 93106, USA}

\begin{abstract}
The theoretical description of non-Fermi liquids is among the most challenging questions in strongly correlated quantum matter.
While there are many experimental candidates for such phases, few examples can be studied using controlled theoretical approaches.
Here we introduce a novel conceptual perspective to analytically describe the non-Fermi liquid physics of gapless spin liquids
arising in a family of two- and three-dimensional Kagome models.
We first discuss a symmetry-protection mechanism that is responsible for the stability of the gapless phase.
We then re-cast the two-dimensional Kagome spin model as a chiral Kondo lattice
model that can be tackled analytically using what is in a sense a generalization of coupled-wire constructions.
We provide extensive numerical evidence for a gapless spin liquid in the two-dimensional Kagome model
and discuss generalizations to three dimensions.
\end{abstract}

\maketitle

% \tableofcontents

%%%%%%%%%%%%%%%%%%%%%%%%%%%%%%%%%%%%%%%%%%%%%%%%%%%%%%%%%%%%%%%%%%

\section{Introduction}

One of the major outstanding challenges of condensed matter theory over the past few decades has been to formulate a fundamental theoretical understanding of non-Fermi liquids -- systems of interacting fermions that evade a description in terms of Landau's Fermi liquid paradigm~\cite{haldane1994,ong2001more}. This ongoing quest 
is fueled by a great number of experimental observations~\cite{lohneysen2007} for various materials that still await an explanation in terms of a comprehensive theory.
The hallmark of a non-Fermi liquid is the emergence of gapless degrees of freedom which cannot be understood as weakly interacting fermionic quasiparticles.
Instead,  these low-energy excitations may be bosons, spinons, fractionalized, or other possibly hitherto unkown degrees of freedom
\footnote{While such non-Fermi liquid physics is fully understood in one spatial dimension in terms of the theory of Luttinger liquids~\cite{Haldane81}, we are solely interested in two or more  spatial dimensions.}.
Probably the best theoretically-established examples of such non-Fermi liquids
are the microscopic lattice models exhibiting ``Bose surfaces'' and ``spinon surfaces'' discussed in the work by
Fisher, Motrunich, Sheng, and collaborators~\cite{Motrunich2007,Sheng2008,Sheng2009,Block2011,Block2011b,Mishmash2011,Jiang2013}. This work has been formative to recognize so-called quantum spin liquids as a natural arena for the occurrence of non-Fermi liquid physics. Quantum spin liquids are non-trivial, often strongly  correlated states of frustrated quantum magnets that evade conventional (symmetry-breaking) order, but instead
retain strong quantum fluctuations
even at zero temperature \cite{balents2010}. A powerful guidance to identify quantum magnets that might harbor spin liquid ground states is given by the Lieb-Schultz-Mattis-Hastings theorem~\cite{Lieb1961,Hastings2004}. It states that the ground state of a translationally invariant system 
of half-integer quantum spins with an odd number of sites in the unit cell
is {\it gapped and degenerate}, or {\it gapless}, in the infinite volume limit (in the absence of spontaneous symmetry breaking). It is precisely this latter scenario which might pave the way for non-Fermi liquid physics in the context of quantum magnetism.

A commonly used framework to describe such low-energy spin liquid physics  
is the notion of fractionalization and emergent gauge fields~\cite{Savary2016}. However, this approach remains uncontrolled for most systems due to the inability to properly
treat  fluctuations of the gauge field. 
For the rare instances where the gauge field remains static -- such as for the celebrated Kitaev model~\cite{Kitaev2006} -- 
the system typically reduces to non-interacting particles, which in turn can be captured by conventional Fermi liquid theory \cite{OBrien2016}.
In this manuscript, we develop an orthogonal perspective to describe non-Fermi liquid physics arising in a frustrated quantum magnet on the Kagome lattice. It is rooted in two seemingly separate approaches, namely (i) the identification of a symmetry protection mechanism and (ii) the reformulation of the spin model in terms of a novel, {\it chiral} Kondo lattice model. The latter is more amenable to an analytical description
than ordinary (i.e. non-chiral)  Kondo lattice models~\cite{Coleman2015}. 
In particular, 
the  analytical treatment of the chiral model 
can be viewed, in some sense, as a generalization of coupled-wire constructions~\cite{Kane2002,KaneTeo2014}. 
This novel perspective on non-Fermi liquid physics, aspects of which were previously developed in Ref.~\onlinecite{bauer2013}, grew out of a conceptually similar treatment of {\it gapped} spin liquids in related Kagome quantum magnets that were shown to harbor a chiral topological
(i.e. gapped)  phase~\cite{Bauer2014}.
There,
unambiguous evidence for the emergence of such a chiral spin liquid with emergent anyon excitations was provided through large-scale numerical simulations based on the density-matrix renormalization group (DMRG). Here we use similar numerical approaches to support our analytical perspective, confirming the gapless nature of the proposed non-Fermi liquid phase.

%%%%%%%%%%%%%%%%%%%%%%%%%%%%%%%%%%%%%%%%%%%%%%%%%%%%%%%%%%%%%%%%%%

\subsection*{Models and line of argument}
\label{SubSectionModelsAndLineOfArgument}

\begin{figure}[t]
  \includegraphics{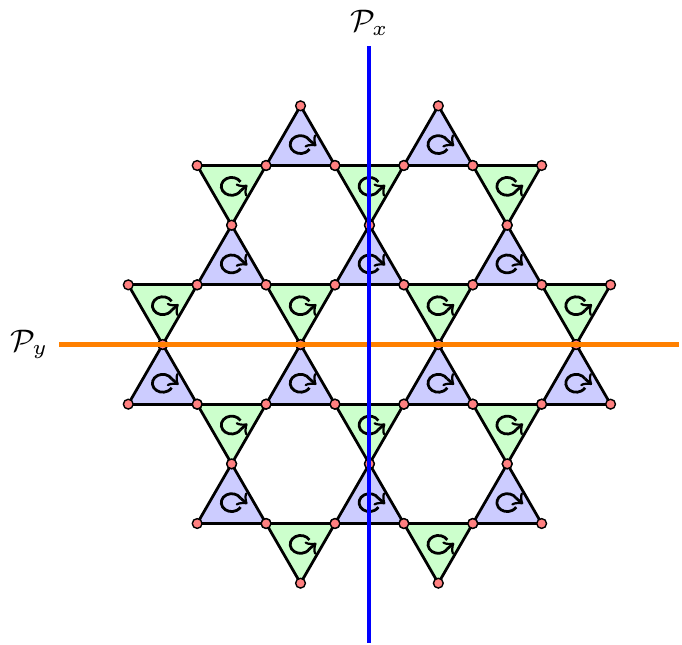}
  \caption{{\bf Staggered Kagome model.}
   		The sign choice of the chiral interactions per elementary triangle is indicated by the arrows and color-coding.
		The two fundamental mirror symmetries of the system, $\mathcal{P}_x$ and $\mathcal{P}_y$, are indicated
		by the solid lines.  \label{fig:stag} }
\end{figure}

The model of our primary interest is given in terms of SU(2) spin-$\frac{1}{2}$ degrees of freedom on the sites of the Kagome lattice and
is described, 
in terms of 
the so-called scalar spin chirality
$\chi_{ijk} = \vec{S}_i \cdot (\vec{S}_j \times \vec{S}_k)$, 
by the SU(2)-invariant Hamiltonian
\begin{equation} 
	\label{eqn:SpinH}
	H = \sum_{i,j,k \in \bigtriangleup} \chi_{ijk} - \sum_{i,j,k \in \bigtriangledown} \chi_{ijk} \,.
\end{equation}
Here, the sites $i$, $j$ and $k$ are always ordered clockwise around the 
two types 
 of elementary triangles
(upwards and downwards oriented)
 $\bigtriangleup$, $\bigtriangledown$ of the Kagome lattice.
The Kagome lattice can be understood as a honeycomb lattice of corner-sharing triangles (formed by their centers - see Fig. \ref{fig:stag}); 
in this language, the Hamiltonian corresponds to a staggered choice of the sign of the coupling to the spin chirality. The choice of chiralities is illustrated in Fig.~\ref{fig:stag}.

\begin{figure}
  \includegraphics{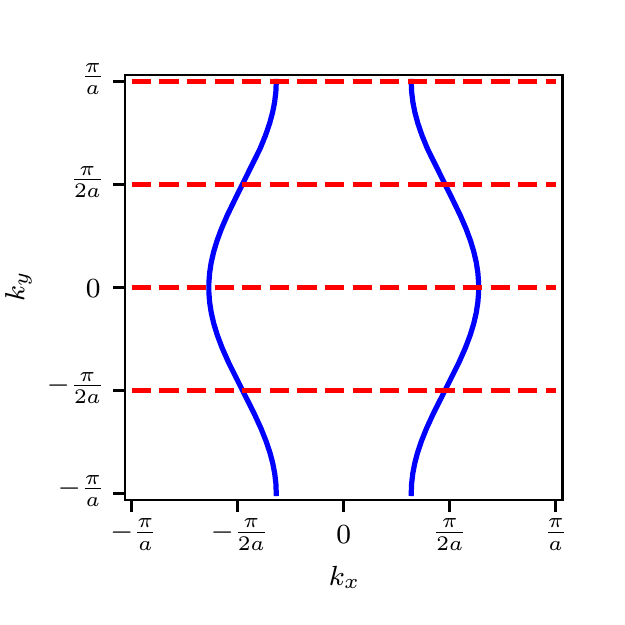}
  \caption{{\bf Spinon surface} of a two-dimensional system. 
  		The dashed red lines indicate the momentum cuts accessible in the quasi-one-dimensional setting 
		of a ladder of width $W=4$.
  		\label{fig:spinonsurface} }
\end{figure}

Our line of argument to demonstrate that this model supports a gapless spin liquid with a spinon surface 
proceeds along the following path: 
One perspective (Section~\ref{sec:quasi1d}) we employ is to look at this model in a {\it quasi-one-dimensional} setting, which makes it analytically more tractable and where we can employ powerful DMRG numerics. Importantly, we note that this dimensional crossover is not obstructing our view on the two-dimensional model as we can make a stringent connection between the one- and two-dimensional limits of its {\em gapless} ground state. In particular, for a 2D system that exhibits a spinon surface, as conceptually depicted in Fig.~\ref{fig:spinonsurface}, the reduction to a quasi-one-dimensional ladder system (in which we make one of the two spatial directions finite, but keep the other spatial direction extended) can be systematically understood \cite{Sheng2008}. The finite ladder width $W$ corresponds to 
%cutting
intersecting the spinon surface $W$ times as illustrated in Fig.~\ref{fig:spinonsurface}, thus resulting in quasi-one-dimensional systems whose number of gapless degrees of freedom (and thus the central charge of the associated conformal field theory) grows linearly with the ladder width $W$. It is precisely this perspective which allows us to use DMRG calculations to study the spin model at hand.

Another perspective (Section~\ref{sec:symmetry}) connects the presence of a surface of gapless
lines to the symmetries of the model, in particular the observation that the Hamiltonian~\eqref{eqn:SpinH}
changes sign under certain lattice reflections.
We rigorously establish 
for the case of a technically simpler, but closely related model
of free Majorana fermions (with a characteristic onsite number of degrees of freedom of $\sqrt{2}>1$),
that these gapless modes are protected by symmetries. 
While certainly being distinct from the interacting spin model,
% it
the Majorana fermion model crucially shares its relevant lattice symmetries.
Replacing the scalar spin chirality $\chi_{ijk}=\vec{S}_i \cdot (\vec{S}_j \times \vec{S}_k)$ in Eqn.~\eqnref{eqn:SpinH} with the quadratic
Majorana fermion term
$\tilde{\chi}_{ijk} = \i (\gamma_i \gamma_j + \gamma_j \gamma_k + \gamma_k \gamma_i)$ (where $\gamma_i = \gamma_i^\dagger$ and
$\lbrace \gamma_i, \gamma_j \rbrace = 2\delta_{ij}$), which like the scalar spin chirality breaks time-reversal and parity symmetry, we obtain
the Majorana Hamiltonian 
\begin{align} 
	\label{eqn:MajH}
	H = \sum_{i,j,k \in \bigtriangleup} \tilde{\chi}_{ijk} - \sum_{i,j,k \in \bigtriangledown} \tilde{\chi}_{ijk}.
\end{align}
where the indices $i$, $j$, $k$ in these sums are again all ordered clockwise around the elementary triangles.
Given the non-interacting nature of this model we can determine its Fermi surface structure both using explicit calculations as well as the symmetry arguments outlined above.
This Majorana model also allows us to explicitly illustrate the above-mentioned dimensional crossover. 
Beyond their common symmetry properties, the Majorana model \eqref{eqn:MajH} and original spin model \eqref{eqn:SpinH} can be connected via a Gutzwiller projection \cite{Gutzwiller1963,Gutzwiller1965} as discussed in Section~\ref{sec:Gutzwiller}.

A third perspective (Section~\ref{sec:weaving})  
is the aforementioned chiral Kondo lattice model.
Starting from decoupled Heisenberg chains the Kagome lattice can be built up by adding interstitial rows of spin-1/2's, each coupled to the adjacent chains via a bowtie of an up- and a down-pointing triangle. 
(See Fig.s 
\ref{fig:zigzag}-\ref{SpinChains3}
below.)
This
% is 
turns out to represent a {\it chiral} version of the Kondo lattice problem, in which the spins couple to only one type of mover (right or left), therefore eliminating back-scattering. This Kondo coupling ends up ``weaving''
% ``weaves" 
the gapless chiral modes, thereby  generating a stack of parallel chiral edge states in real space, which when viewed from 
momentum space correspond to the sought-after spinon surfaces. We 
% check 
confirm the validity of this weaving argument by explicitly reconstructing the Fermi surface of the Majorana fermion model using this construction.

Taken together, these three complementary perspectives provide us with a comprehensive argument supporting
the emergence of spinon surfaces in  Kagome spin models. This approach can be generalized to other Kagome-like lattice structures. Of particular interest might be three-dimensional lattices of corner-sharing triangles (such as the hyperkagome lattice) that along this line of arguments are proposed to harbor spin liquids with (two-dimensional) spinon surfaces (Section~\ref{sec:3d}).

%%%%%%%%%%%%%%%%%%%%%%%%%%%%%%%%%%%%%%%%%%%%%%%%%%%%%%%%%%%%%%%%%%

\section{Symmetry protected metals}
\label{sec:symmetry}

In this section, we discuss Hamiltonians $H$ that anticommute with a symmetry operation $\Ss$ of the lattice, i.e.
\begin{equation} \label{eqn:Hanticomm}
\lbrace H, \Ss \rbrace = 0.
\end{equation}
We will examine under what conditions such an ``antisymmetry'' guarantees that a model exhibits gapless excitations,
focusing on the case of non-interacting translationally invariant fermionic systems. We find two suitable conditions:
first, if the system has an odd number of sites per unit cell, such an antisymmetry guarantees that at least one energy
band vanishes on the submanifold of momentum space that is invariant under the symmetry.
In the case where the number of sites per unit cell is even, we find that gapless modes are guaranteed
if the symmetry acts on the fermion operators at invariant momenta as an operator with finite signature,
i.e. possessing  a different number of positive and negative eigenvalues.

Our results complement previous work on the symmetry stabilization of nodal lines in the literature. In particular, 
it has been found that combinations of inversion and time-reversal symmetry $(PT)$ \cite{Kim2015,Fang2015,Bieri2015,Bieri2016,Chan2016}, sublattice/chiral symmetry \cite{Zhao2013}, crystal reflection \cite{Schnyder2014}, non-centrosymmetric  \cite{Yamakage2016} and non-symmorphic lattice symmetries \cite{Yang2017} 
%as 
are permissible symmetry sets that allow to protect nodal lines.

\subsection{General statement}

\subsubsection{Odd number of bands}
\label{sec:symm_odd}

A particularly simple argument can be made in the case where the number of sites in the unit cell, or equivalently
the number of bands, is odd. To illustrate the argument, consider a fermionic Hamiltonian given by
\begin{equation} \label{eqn:Ham}
H = \sum H_{ij} c_i^\dagger c_j.
\end{equation}
Hermiticity requires that $H_{ij} = H^*_{ji}$. Assuming translational invariance, this can be brought into the form
\begin{equation}
H = \sum_{\sigma \vec{k}} \epsilon_\sigma (\vec{k}) c_{\sigma \vec{k}}^\dagger c_{\sigma \vec{k}},
\end{equation}
where now $\vec{k}$ denotes the lattice momentum and the index $\sigma$ runs over the number of bands in the model.
We consider the effect of a symmetry
$\mathcal{S}$ with $\mathcal{S}^2=\id$, i.e. the symmetry group is $\mathbb{Z}_2$. Crucially, we consider Hamiltonians satisfying
$\mathcal{S} H \mathcal{S} = -H$, and the symmetry is a lattice symmetry that transforms momentum $\vec{k}$ into
$\check{\vec{k}}=f(\vec{k})$, i.e.
$\mathcal{S} c_{\sigma \vec{k}} \mathcal{S} = c_{\check{\sigma} \check{\vec{k}}}$, where $f=f^{-1}$ is an invertible vector-valued function on the Brillouin zone and ${\check{\sigma}}$ denotes a permutation of the band index.
We denote the set of points invariant
under $f$ as $\mathfrak{S}$, i.e. for all $\vec{k} \in \mathfrak{S}$, $f(\vec{k})=\vec{k}$.

Then,
\begin{align}
\mathcal{S} H \mathcal{S} &= \sum_{\sigma \vec{k}} \epsilon_\sigma(\vec{k}) c_{\check{\sigma} \check{\vec{k}}}^\dagger c_{\check{\sigma} \check{\vec{k}}} \\
&= \sum_{\sigma \vec{k}} \epsilon_{\check{\sigma}}(\check{\vec{k}}) c_{\sigma \vec{k}}^\dagger c_{\sigma \vec{k}} \\
&= -H.
\end{align}
To satisfy this, we must have that for each band $\sigma$, there is a band $\check{\sigma}$ such that
\begin{equation}
\epsilon_{\check{\sigma}}(\check{\vec{k}}) = -\epsilon_\sigma(\vec{k}).
\end{equation}
In other words, we have seen that the energy bands of $H$ must form a representation of $\mathbb{Z}_2$. A given
band in the spectrum of $H$ can thus either transform into itself under $\mathcal{S}$, or into exactly one
other band. We see that if the number of bands is odd, there must be at least one band $\epsilon_0$ that transforms
into itself under $\mathcal{S}$, $\epsilon_0(\check{\vec{k}}) = -\epsilon_0(\vec{k})$. This implies that this band vanishes on $\mathfrak{S}$,
i.e.
\begin{equation}
\epsilon_0(\vec{k}) = 0\mathrm{\ \ for\ all\ \ }\vec{k} \in \mathfrak{S}.
\end{equation}
If the total number of bands is even, no such constraint arises.

Examples of the structure discussed above naturally arise in the context of Majorana models.
If we require that $H_{ij} = i A_{ij}$, where $A$ is an antisymmetric, real matrix,
we see that we can introduce Majorana fermions $\gamma_i = \gamma_i^\dagger$, $\eta_i=\eta_i^\dagger$ with the standard
anticommutation rules $\lbrace \gamma_i, \gamma_j \rbrace = \lbrace \eta_i, \eta_j \rbrace = 2 \delta_{ij}$, $\lbrace \eta_i, \gamma_j \rbrace = 0$.
In terms of these, the original fermion operators can be expressed as $c_i = (\gamma_i + i \eta_i)/\sqrt{2}$
such that the Hamiltonian~\eqnref{eqn:Ham} becomes
\begin{equation}
\label{LabelEqHamiltonianTwoMajoranaSpecies}
H = i \sum_{ij} A_{ij} \left( \gamma_i \gamma_j + \eta_i \eta_j \right).
\end{equation}
In this sense, we can interpret the single-particle spectrum of $H_{ij} = i A_{ij}$ as the spectrum of the Majorana fermions $\gamma_i$ with Hamiltonian $H_\gamma = i \sum A_{ij} \gamma_i \gamma_j$.

For illustration purposes, let us consider a translationally invariant one-band system of Majorana fermions,
\begin{equation}
H = i \sum_{\vec{x},\vec{y}} A(\vec{x}-\vec{y}) \gamma_{\vec{x}} \gamma_{\vec{y}},
\end{equation}
where $\vec{x}$ and $\vec{y}$ denote points on a real-space lattice.
Upon Fourier expansion $\gamma_{\vec{k}} = \sum e^{i \vec{k} \cdot \vec{x}} \gamma_{\vec{x}}$, we obtain
\begin{align}
H &= \sum_{\vec{k}} \epsilon(\vec{k}) \gamma_{\vec{k}} \gamma_{-\vec{k}} &\epsilon(\vec{k}) &= i \sum_{\vec{d}} A(\vec{d}) e^{i \vec{k} \cdot \vec{d}}
\end{align}
Since hermiticity of $H$ requires $A(-\vec{d}) = -A(\vec{d})$, we immediately find that $\epsilon(-\vec{k}) = -\epsilon(\vec{k})$ and thus
\begin{equation}
\epsilon(\vec{k}=0) = 0.
\end{equation}
This gapless point is protected by the inversion symmetry $\mathcal{I} : \vec{x} \rightarrow -\vec{x}$, which always has to
anticommute with the Hamiltonian for such a single-band Majorana model.

A simple example where this is realized is a chain of Majorana fermions, $H=i \sum_i \gamma_i \gamma_{i+1}$, which has a single
band $\epsilon(k) = \sin(ka)$ that vanishes at $k=0$.

%%%%%%%%%%%%%%%%%%%%%%%%%%%%%%%%%%%%%%%%%%%%%%%%%%%%%%%%%%%%%%%%%%

\subsubsection{Generalization including 
even and odd numbers of bands}
\label{sec:symm_even}

In the following, we discuss a more involved line of argument that is applicable to systems with any number of sites in the unit cell. We will show
that a spatial symmetry that anticommutes with the Hamiltonian, and whose signature -- the number of positive minus
the number of negative eigenvalues -- is nonzero, guarantees gapless excitations at the momenta invariant under the symmetry.

We consider translationally-invariant systems on a $d$-dimensional lattice described in terms of Majorana operators $\gamma_\alpha(R)$, where $R$ denotes the origin of a unit cell,
and $\alpha$ a site within the unit cell. We denote as $\vec{s}_\alpha$ the position of the site with respect to the origin of the unit cell.
The operator $\gamma_\alpha(R)$ therefore corresponds to a Majorana at position $R+\vec{s}_\alpha$. The Hamiltonian can be written in the form
\begin{equation}
H = \sum_{\vec{R}_1,\vec{R}_2,\alpha,\beta} h_{\alpha \beta}(\vec{R}_1-\vec{R}_2)\ i \gamma_\alpha(\vec{R}_1) \gamma_\beta(\vec{R}_2).
\end{equation}
In momentum space, we
write
\begin{equation} \label{eqn:kansatz}
\gamma_\alpha (\vec{k}) = \sum_{\vec{R}} e^{i \vec{k} \cdot (\vec{R}+\vec{s}_\alpha)} \gamma_\alpha(\vec{R}),
\end{equation}
to obtain
\begin{eqnarray} \label{eqn:hk}
H &=& \sum_{\alpha,\beta,\vec{k}} h_{\alpha \beta}(\vec{k}) \gamma_\alpha(\vec{k}) \gamma_\beta(-\vec{k}) \\
h_{\alpha \beta}(\vec{k}) &=& e^{i \vec{k} \cdot (\vec{s}_\beta-\vec{s}_\alpha)} \sum_{\vec{dR}} h_{\alpha \beta}(\vec{dR}),
\end{eqnarray}
where we have introduced $\vec{dR}=\vec{R}_2-\vec{R}_1$.
We now consider a symmetry operation $\Ss: \vec{x} \rightarrow \Ss(\vec{x})$ of the form $\Ss(\vec{x}) = \check{\vec{x}}+\vec{v}$ with the property that $\Ss^2=\nu \id$, where
$\nu \in \mathbb{C}$ and $\nu \neq 0$, and ${\vec x} \to \check{\vec{x}}$ denotes a linear operator.
For an example, consider in one dimension  a reflection around $x=1/2$, i.e. $\Ss(x) = -x+1$; then $\check{x}=-x$ and $v=1$.
Let the action of $\Ss$ on the Majorana operators in real space be
\begin{equation}
\Ss \gamma_\alpha(\vec{R}) \Ss = \gamma_{\check{\alpha}}(\check{\vec{R}}+\vec{o}_\alpha),
\end{equation}
where $\vec{o}_\alpha \in \mathbb{Z}^{\otimes d}$ is an offset that accounts for the fact that the new operator may be found in a different unit cell, and in slight abuse of notation we denote
by $\check{\alpha}$ the index of the site within the unit cell that the $\alpha$'th site is mapped to under the symmetry.
Then, we find that the action of $\Ss$ on the momentum-space operators $\gamma_\alpha(\vec{k})$ is given by
\begin{align} \label{eqn:gammasymmaction}
\Ss \gamma_\alpha(\vec{k}) \Ss = e^{i \check{\vec{k}} \cdot (\check{\vec{s}}_\alpha-\vec{o}_\alpha-\vec{s}_{\check{\alpha}})} \gamma_{\check{\alpha}}(\check{\vec{k}})
\end{align}
For details of the derivation, see App.~\ref{app:hyperkagome}.
If we let $\psi(\vec{k}) = \left( \gamma_0(\vec{k}), \ldots, \gamma_N(\vec{k}) \right)^T$, then we can define a matrix $U(\vec{k})$ such that $\Ss \psi(\vec{k}) \Ss = U(\vec{k}) \psi(\check{\vec{k}})$.
Focusing on momenta $\vec{k}^*$ that are invariant under the symmetry, i.e. $\vec{k}^* = \check{\vec{k}^*}$, the condition that
$\Ss^2 = \nu \id$ implies that also $U(\vec{k}^*)^2=\nu \id$, 
and thus the eigenvalues of $U$, which we denote
by 
%as 
$\lambda_j$, must satisfy $\lambda_j = \pm \sqrt{\nu}$ (all eigenvalues have equal magnitude, but may differ in sign).

We now use the condition of Eqn.~\eqnref{eqn:Hanticomm}, namely that $\Ss H \Ss = -H$, and consider invariant momenta $\vec{k}^*$.
Using as shorthand notation $h^* = h(\vec{k}^*)$ and $U^* = U(\vec{k}^*)$, we insert the explicit form of the Hamiltonian~\eqnref{eqn:hk} to find
\begin{equation}
U^* h^* = - h^* U^*
\end{equation}

Let $v$ be an eigenvector of $U^*$ with eigenvalues $\lambda = \pm \sqrt{\nu}$. Then, anticommutativity of $U^*$ and $h^*$ implies that
$U^* h^* v = -h^* U^* v = -\lambda h^* v$, i.e. $h^* v$ is either an eigenvector of $U^*$ with eigenvalue $-\lambda$ or $v$ is an eigenvector of $h^*$ with eigenvalues 0. We can thus use
$h^*$ to construct a map between the $+\sqrt{\nu}$ and $-\sqrt{\nu}$ eigenspaces of $U^*$. Now consider the case where the signature of $U^*$ is non-vanishing, i.e. the dimensions
of the $+\sqrt{\nu}$ and $-\sqrt{\nu}$ eigenspaces of $U^*$ are not equal. It then follows immediately that the kernel of $h^*$ is non-empty, i.e. $h^*$ has vanishing eigenvalues.

To summarize, we have shown a more stringent condition for
cases where a Hamiltonian that anticommutes with a spatial symmetry $\Ss$ exhibits
gapless modes. Considering the operator representation of this symmetry on momentum-space fermion operators at momenta that are left invariant
by the symmetry, we find that gapless modes must exist if the operator has an unequal number of positive and negative eigenvalues, i.e. non-zero
signature. Clearly, in the case of an odd number of bands, this condition collapses to the simpler one discussed in Sec.~\ref{sec:symm_odd}, since
an odd-dimensional matrix must necessarily have non-zero signature. We will use the more
general argument to argue for gapless modes in quasi-one-dimensional
realizations of our model below, and to argue for relevant three-dimensional cases in Sec.~\ref{sec:3d}.

%%%%%%%%%%%%%%%%%%%%%%%%%%%%%%%%%%%%%%%%%%%%%%%%%%%%%%%%%%%%%%%%%%

\subsection{Applications to Kagome systems}
\label{SubSectionKagomeModel}

\subsubsection{Two-dimensional case}

As a first step to illustrating the role of the symmetry-protection mechanism in the context of the models discussed throughout this paper,
we turn to the model of Majorana fermions (first introduced in Eqn.~\eqnref{eqn:MajH}) given by
\begin{align} \label{eqn:MajH2}
H = \sum_{i,j,k \in \bigtriangleup} \tilde{\chi}_{ijk} - \sum_{i,j,k \in \bigtriangledown} \tilde{\chi}_{ijk}.
\end{align}
where the $i$, $j$, $k$ in these sums are all ordered counter-clockwise around the triangles, and
$\tilde{\chi}_{ijk} = \i (\gamma(i) \gamma(j) + \gamma(j) \gamma(k) + \gamma(k) \gamma(i))$. This can be interpreted as a Majorana model
$H=\i \sum A_{ij} \gamma(i) \gamma(j)$ with $A_{ij} \in \lbrace 0, \pm 1 \rbrace$, and the sign as indicated by the arrows in Fig.~\ref{fig:stag}.
There are three sites in a unit cell.

\begin{figure}
  \centering
  (a) \\
  \includegraphics[width=0.7\columnwidth]{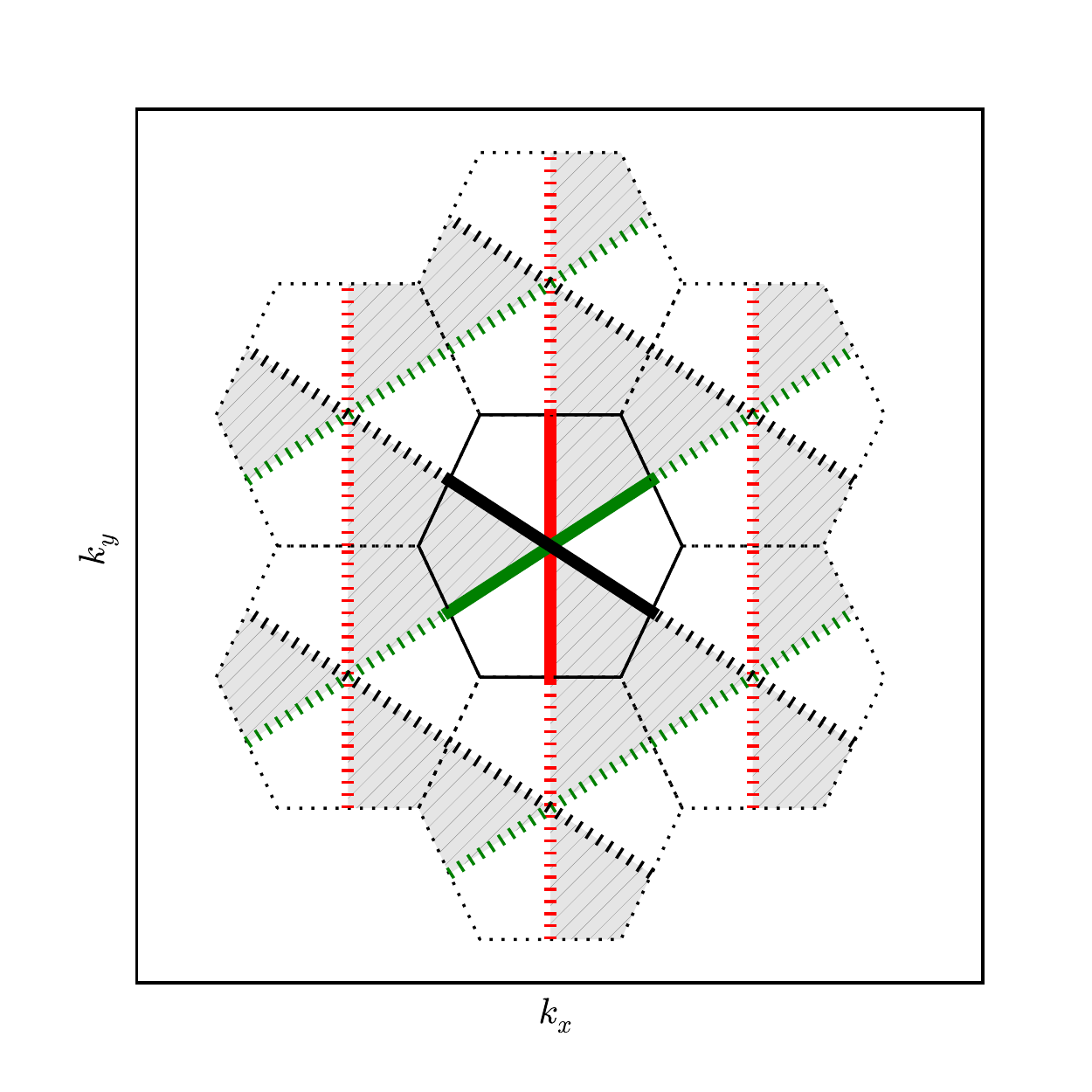} \\
  (b) \\
  \includegraphics[width=0.3\columnwidth]{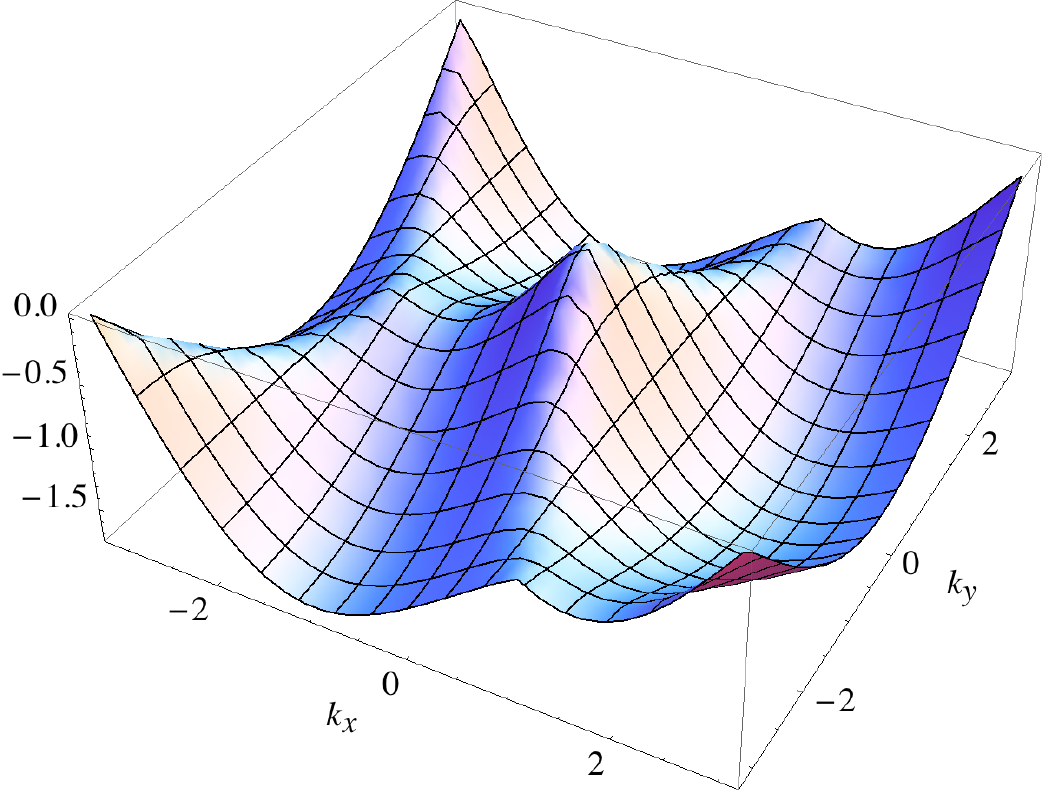}
  \includegraphics[width=0.3\columnwidth]{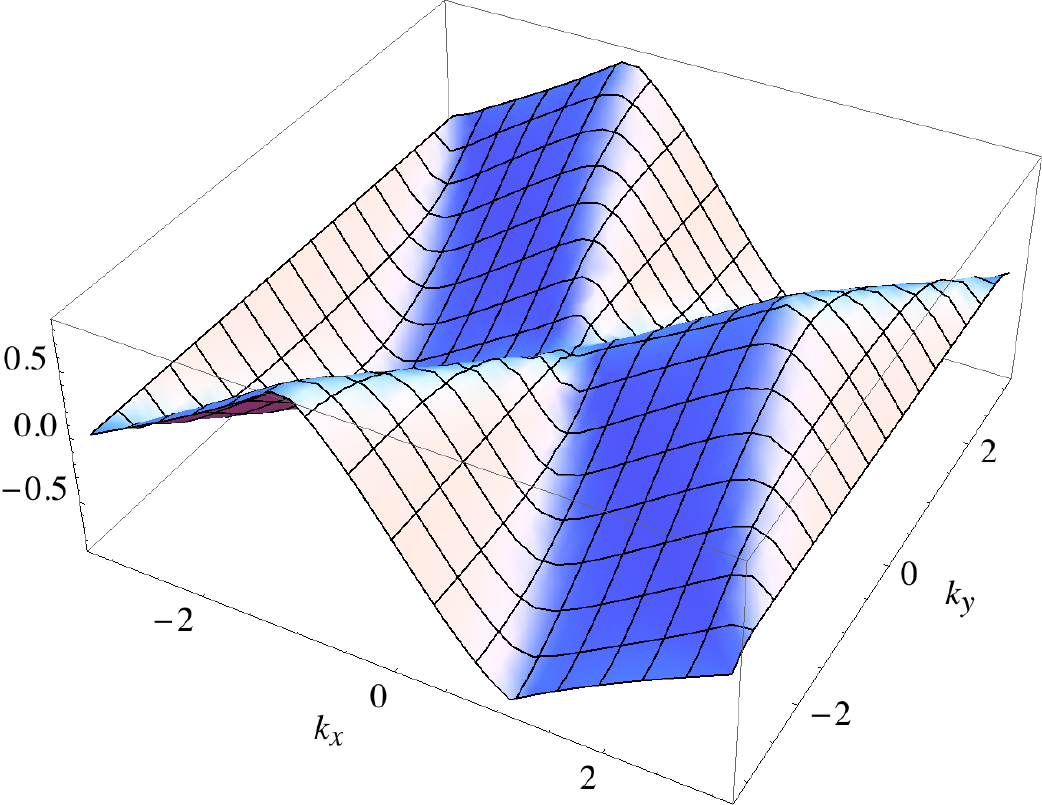}
  \includegraphics[width=0.3\columnwidth]{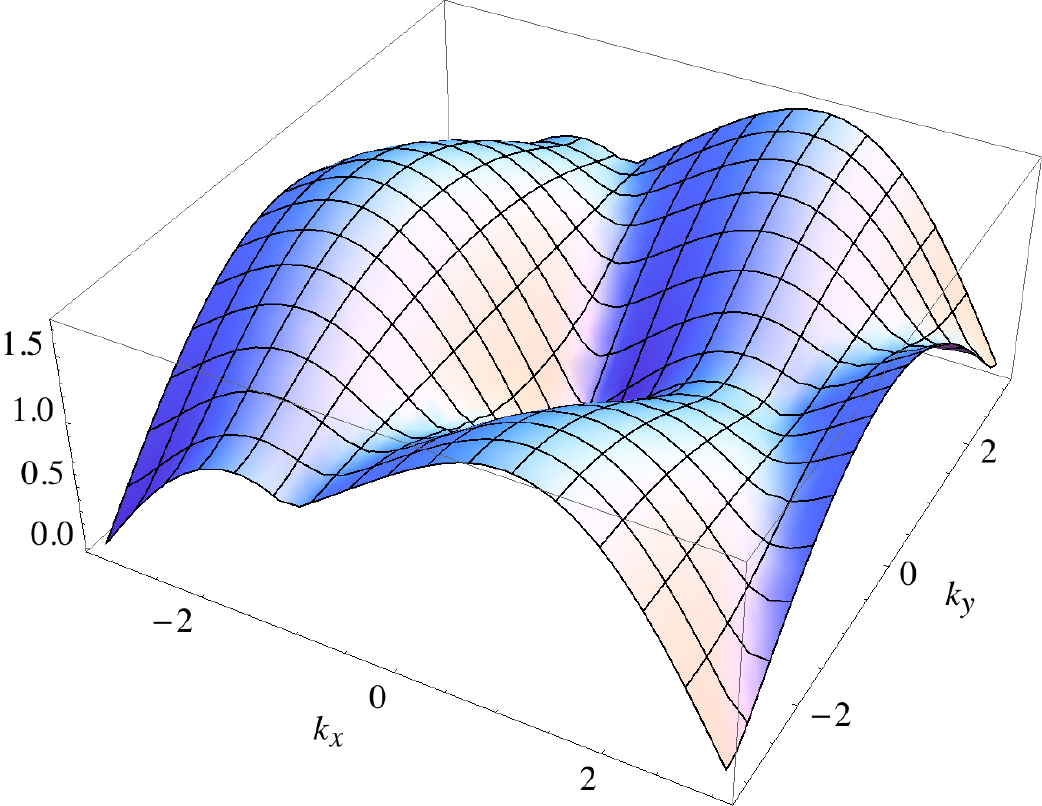}
  \caption{{\bf Band structure} with three gapless lines indicated by the red/green/black lines in panel (a)
  		and a three-dimensional illustration in panel (b).  \label{fig:bands} }
\end{figure}

For our considerations here,
the relevant symmetries are a three-fold rotation symmetry $\mathcal{R}_3$ around the center of a hexagon
and two mirror (reflection) symmetries about the lines  labeled $\mathcal{P}_x$
and $\mathcal{P}_y$ in Fig.~\ref{fig:stag}. The system is even under the rotation, $\mathcal{R}_3 H \mathcal{R}_3 = H$, and behaves
as follows under the mirror symmetries:
\begin{align}
\mathcal{P}_y H \mathcal{P}_y &= H & \mathcal{P}_x H \mathcal{P}_x &= -H.
\end{align}
Since the product of the two mirror symmetries 
yields inversion, $\mathcal{P}_x \mathcal{P}_y = \mathcal{I}$, the system is odd under
inversion, as well as under all three distinct rotated versions of the $\mathcal{P}_x$ mirror symmetry $\mathcal{P}_x \mathcal{R}_3^n$,
with $n=0,1,2$. The mirror symmetry $\mathcal{P}_x$ acts in momentum space as $\vec{k}=(k_x,k_y) \rightarrow (-k_x,k_y)$,
and correspondingly for the rotated versions. Noting that the Kagome lattice has three sites per unit cell and thus an odd number of
bands, we find that there must be at least one band $\epsilon(k_x,k_y)$ which satisfies for all $t \in \mathbb{R}$
\begin{subequations} \label{eqn:fs} \begin{gather}
\epsilon(0,t) = 0 \\
\epsilon(\cos(2\pi/3)t,\sin(2\pi/3)t) = 0 \\
\epsilon(\cos(4\pi/3)t,\sin(4\pi/3)t) = 0.
\end{gather} \end{subequations}
This gives rise to the band structure shown in Fig.~\ref{fig:bands}(a) with three gapless lines that meet at $\vec{k}=0$.
This result is confirmed by solving the model exactly to find the three
bands shown in Fig.~\ref{fig:bands}(b). Here, we find that the top and bottom bands satisfy $\epsilon_1 \geq 0$, $\epsilon_3 \leq 0$,
respectively, and they transform into each other under the odd mirror symmetries, while $\epsilon_2$ transforms into itself
and satisfies Eq.~\eqnref{eqn:fs}, thus giving rise to the Fermi surface.
The shading in Fig.~\ref{fig:bands}(a) shows the regions of the Brillouin zone with negative and positive single-particle
energies in the band $\epsilon_2$, which correspond to filled and empty regions when formulated in terms of
 complex fermions
[as in the paragraph above Eq.~\eqref{LabelEqHamiltonianTwoMajoranaSpecies}].
A Fermi surface of this shape was found for a closely related model on the triangular lattice in Ref.~\onlinecite{biswas2011},
where it was used as mean-field ansatz for the parton construction of a spin liquid.

\subsubsection{Quasi-one-dimensional realizations}
\label{sec:ff1d}

\begin{figure}[b]
  \centering
  \includegraphics{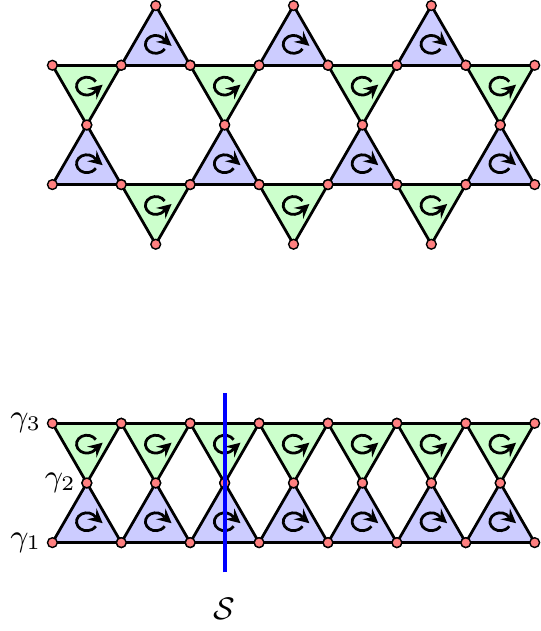}
  \caption{{\bf Kagome cylinders} of  width $W=2$ (upper panel - the lattice site in the op and bottom row are identified), and 
  		its representation as a quasi-one-dimensional ladder system (lower panel - see text).
		\label{fig:zigzag} }
\end{figure}

We now move on to consider quasi-one-dimensional realizations of the model~\eqnref{eqn:MajH2}. More specifically,
we consider Kagome cylinders of width $W=2$ (``2-leg Kagome cylinders''),
as shown in the top panel of Fig.~\ref{fig:zigzag}. These can be represented as quasi-one-dimensional systems with a three-site unit cell, shown in
the lower panel of Fig.~\ref{fig:zigzag}. The Hamiltonian on this lattice takes the form
\begin{multline}
H = i \sum_i \left[ \gamma_1(i) \gamma_2(i) + \gamma_2(i) \gamma_1(i+1) + \gamma_1(i+1) \gamma_1(i) \right. \\
 \left. + \gamma_3(i) \gamma_2(i) + \gamma_2(i) \gamma_3(i+1) + \gamma_3(i+1) \gamma_3(i) \right].
\end{multline}
For this choice of boundary conditions,
vertical reflections about the horizontal center axis are a symmetry of the system. We can thus rewrite the Hamiltonian in terms of
degrees of freedom that are even and odd under this symmetry:
\begin{align}
\text{Even:} &\ \ \ \gamma_+(i) = \gamma_1(i) + \gamma_3(i), \ \ \gamma_2(i) \\
\text{Odd:} &\ \ \ \gamma_-(i) = \gamma_1(i) - \gamma_3(i).
\end{align}
The Hamiltonian decouples into the even and odd sectors:
\begin{align}
H &= H_e \oplus H_o \\
H_e &= \!\begin{multlined}[t]
	i \sum_i \left[ \gamma_+(i) \gamma_2(i) + \gamma_2(i) \gamma_+(i+1) \right . \nonumber \\
	\left. + \frac{\gamma_+(i+1) \gamma_+(i)}{2} \right]
   \end{multlined} \\
H_o &= i \sum_i \frac{1}{2} \gamma_-(i+1) \gamma_-(i) \nonumber
\end{align}
From inspection of the lower panel of Fig.~\ref{fig:zigzag}, it is also apparent that the Hamiltonian is odd under a 
horizontal reflections about an axis that passes through the sites on sublattice 2, as denoted by $\mathcal{S}$ in Fig.~\ref{fig:zigzag}.
Since this symmetry commutes with the
other reflection symmetry, it is a symmetry of each sector separately.
For the odd sector, which has only a single band, this implies that this band must
be gapless at the invariant momentum $k=0$ per the simple arguments of Sec.~\ref{sec:symm_odd}.

The even sector has two bands, and therefore the arguments of Sec.~\ref{sec:symm_even} need to be applied to understand
whether it hosts gapless modes at $k=0$. The unit cell of the system has two sites with operators $\gamma_+(i)$ at $s_+=0$,
and operators $\gamma_2(i)$ at $s_2 = 1/2$. The symmetry operation is a reflection about the $\gamma_2$ operator
in the unit cell at $R=0$: $\Ss(x) = -x+1$ (and thus $\check{x} = -x$). Then we have that
\begin{align}
\Ss \gamma_+(R) \Ss &= \gamma_+(-R+1) & \Ss \gamma_2(R) \Ss &= \gamma_2(-R)
\end{align}
To see the second case, consider that $\gamma_2(R)$ is at position $R+s_1 = R+1/2$, which under $\Ss$ goes to $-R+1/2$. The offset vectors are then
$o_+ = 1$ and $o_2 = 0$. For $\gamma_+$, using $\check{s}_+=0$, $s_{\check{+}} = 0$, we find the momentum-space operators to transform as 
\begin{equation}
\Ss \gamma_+(k) \Ss = e^{i (-k) (0-1-0)} \gamma_+(-k) = e^{ik} \gamma_+(-k).
\end{equation}
Similarly, using $\check{s}_2 = -1/2$, $s_{\check{2}} = 1/2$, and $o_2 = 0$, we have
\begin{equation}
\Ss \gamma_2(k) \Ss = e^{i (-k) (-1/2 - 0 -1/2)} \gamma_2(-k) = e^{ik} \gamma_2(-k).
\end{equation}
If we now consider a momentum that is invariant under the symmetry, $k=-k=0$, we find that $U = \id$ and thus has signature $+2$, implying that the Hamiltonian
vanishes at $k=0$.
For illustration purposes, we consider in Appendix~\ref{app:hyperkagome} the case of a bond- rather than site-centered inversion, which turns out
not to guarantee any gapless modes.

We therefore find that the quasi-one-dimensional realization shown in Fig.~\ref{fig:clusters} has at least two gapless modes, one in each
of the odd and even sector. Exactly solving the model shows that the number of gapless modes is exactly two. Generalizing this result
to wider cylinders, one finds that the number of gapless points guaranteed by symmetry scales linearly in the width of the cylinder.
As laid out in the Introduction, such a linear scaling of the number of gapless modes with the width of quasi-one-dimensional ladders
is a precursor of the Fermi surface recovered in the two-dimensional limit. 
While for the system of non-interacting fermions, we can tackle the two-dimensional system directly, such a quasi-one-dimensional approach
will be crucial for studying the more challenging case of interacting spins.

\section{Chiral Kondo lattice model}
\label{sec:weaving}

%%%%%%%%%%%%%%%%%%%%%%%%%%%%%%%%%%%%%%%%%%%%%%%%%%%%%%%%%

\begin{figure}[t]
  \includegraphics{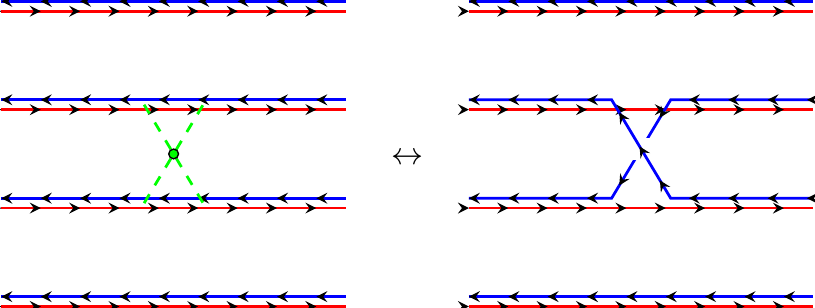}
  \caption{{\bf Bowtie realization of the 2-channel Kondo effect} for a system of counterpropagating 
  		one-dimensional edge states (indicated by the red/blue lines) coupled via an interstitial site (green dot). }
  \label{SpinChains1}
\end{figure}

The idea of the chiral Kondo lattice model is to construct the Kagome spin model \eqref{eqn:SpinH} by starting from a well-understood gapless system: a stack of parallel, but decoupled Heisenberg chains. These gapless Heisenberg chains can each be formulated in terms of right-moving and left-moving spin degrees of freedom. It is precisely these gapless modes of two neighboring chains
that we couple to an interstitial impurity spin, located between the two chains (see the left panel of Fig.~\ref{SpinChains1}).
Owing to the chirality of the coupling (mediated along a bowtie of an up and down pointing triangle), this 
turns out to give rise to a Kondo effect that
does  not affect the right-moving degrees of freedom of either chain, which therefore keep propagating unimpededly along each spin chain. In contrast, the left-moving gapless degrees of freedom of the two chains
 couple to the impurity spin $1/2$ and in fact end up  interchanging (``weaving'')  as illustrated in the right panel of Fig.~\ref{SpinChains1}. 
 This situation 
 realizes a chiral $2$-channel Kondo effect \cite{LudwigAffleck1991,AffleckLudwig1991}. (See Sect. \ref{sec:technical}
for more details.)

\begin{figure}[t]
  \includegraphics{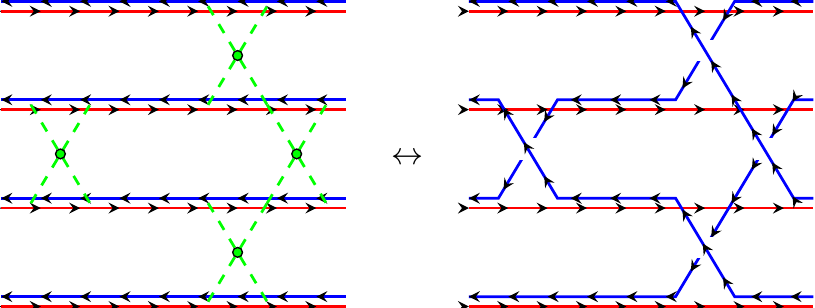}
  \caption{{\bf Multiple bowties} upon the coupling of several edge states.}
  \label{SpinChains2}
\end{figure}

\begin{figure}[t]
  \includegraphics{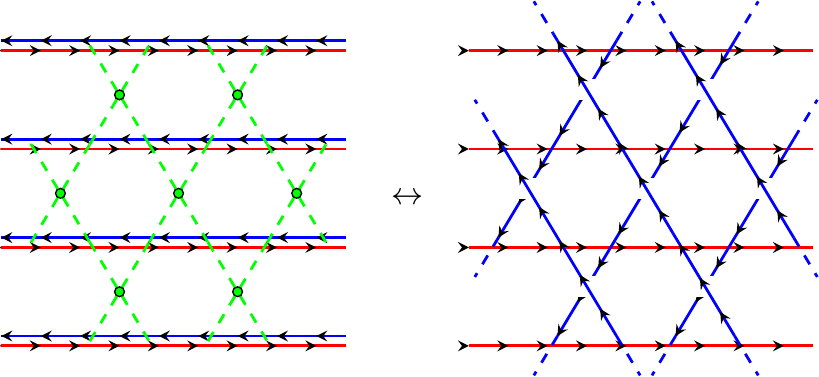}
  \caption{{\bf Chiral Kondo lattice model} arising from coupling all edge modes via the interstitial sites of the Kagome lattice. }
  \label{SpinChains3}
\end{figure}

Given this basic building block of
 the $2$-channel Kondo effect 
arising from the inclusion of a single
 interstitial spin into 
our construction, we can imagine adding further interstitial spins one-by-one to eventually obtain the entire Kagome lattice. It is important to note that because of the chirality there is no backscattering
allowing us to add the interstitial spins independently of each other.
In terms of the gapless modes, each realization of the $2$-channel Kondo effect amounts to the aforementioned weaving of modes, which results in the sequence of weavings illustrated in Figs.~\ref{SpinChains1}, \ref{SpinChains2} and \ref{SpinChains3}. 
The final outcome after including all interstitial spins are three stacks of parallel running lines of {\it chiral}  gapless modes as illustrated in the right panel of Fig.~\ref{SpinChains3}. When viewed in momentum space these modes give precisely rise to a spinon surface
at the same location in momentum space as that of the Fermi surface (line)
% analogous to the one
 shown in Fig.~\ref{fig:bands}.

%%%%%%%%%%%%%%%%%%%%%%%%%%%%%%%%%%%%%%%%%%%%%%%%%%%%%%%%%

\subsection{Technical implementation}
\label{sec:technical}

We begin by reviewing briefly the low-energy description of a single spin-$1/2$ chain. The lattice spin operator $\vec{S_n}$ at lattice site $n$ is represented in the low-energy
continuum field theory as a combination of left- and right-moving degrees of freedom,
\begin{equation} \begin{split}
\label{Continuum} a^{-1} \vec{S}_n \rightarrow \vec{J}_L(z_n) + \vec{J}_R(\bar{z}_n) + \\ i(-1)^n A \times \mbox{tr}(g(z_n,\bar{z}_n)\vec{\sigma}).
\end{split} \end{equation}
The holomorphic (``left-moving'') and anti-holomorphic (``right-moving'') coordinates $z_n$ and $\bar{z}_n$ correspond to the space and imaginary time coordinates $z_n=\tau+ix_n$ and $\bar{z}_n=\tau-ix_n$ where $x_n=na$
and $a$ is the lattice spacing.
Furthermore, $g$ is an $SU(2)$ matrix field described by the $SU(2)_1$ Wess-Zumino-Witten [WZW] action governing the spin degree of freedom, $\vec{J}_{L,R}$ are Kac-Moody currents of the $SU(2)_1$ WZW conformal field theory. The constant $A$ is non-universal.

A spin-$1/2$ impurity $\vec{S}_\text{imp}$ is now placed between two chains and coupled to the spins on site $n$ and $n+1$ of either chain
through 
a  staggered scalar spin chirality interaction, as shown in Fig.~\ref{SpinChains1}. (See also the top panel of
Fig. \ref{fig:zigzag}.)
Explicitly, denoting the spins on the upper chain 
by  $\vec{S}$ and the spins on the lower chain 
by  $\vec{T}$ as well as the coupling constant
by $K$,
this reads
\be
\label{SpinForm2channel} -K S^a_\text{imp} \epsilon^{abc}(S^b_n S^c_{n+1}+T^b_n T^c_{n+1}).
\ee
Without loss of generality we choose $K>0$. (For $K<0$ we need to exchange the role of right- and left-movers in the discussion below.)

Let us focus momentarily only on the
coupling of the spin-1/2 impurity to the spins ${\vec S}_n$ of the upper chain.
When expressed in terms of the long-wavelength degrees of freedom of Eqn.~\eqnref{Continuum}, we find
(see appendix \ref{app:ChiralKondo} for details of the calculation)
\be
\label{CurrentForm2channel} -K S^a_\text{imp} \epsilon^{abc} S^b_n S^c_{n+1} \rightarrow - K S^a_\text{imp} (J^a_R(\bar{z}_n)-J^a_L(z_n)).
\ee
Crucially, there is a sign difference between the coupling of the left-moving and right-moving spin currents to the impurity.
This
is expected based on symmetry considerations alone: since the microscopic Hamiltonian is odd under reflection $\mathcal{P}_x$ (see Fig.
\ref{fig:stag}), this coupling of the low-energy degrees of freedom also
changes sign under reflection, which in turn  exchanges left- and right-movers.

This coupling to the impurity spin is precisely the coupling that occurs in the Kondo effect. In the Kondo effect the coupling to the impurity is antiferromagnetic (positive); for opposite sign of the coupling (ferromagnetic/negative) the impurity spin simply decouples
at low energy. Here the coupling of the right-moving (left-moving) current with the impurity spin
 is ferromagnetic (antiferromagnetic). Since the Kondo coupling is known to be marginally irrelevant (in the renormalization group sense) for negative coupling, and marginally relevant for positive coupling, the coupling of the right-movers to the impurity will flow to zero at long scales
under the renormalization group. Therefore only the coupling between the left-movers and the impurity survives 
at low energy and
 the impurity only couples to the left-moving gapless spin degrees of freedom. This confirms that only the left-moving currents will undergo a Kondo effect with the impurity while each right-moving current of the spin $1/2$ chains will remain unaffected.

Returning to Eqn.~\eqnref{SpinForm2channel}, we see that the impurity spin couples 
with equal strength
 to the $\vec{S}$ and $\vec{T}$ spins from each adjacent chain. Following the
arguments given above, the impurity spin effectively couples to the sum of the two left-moving spin current densities.
We therefore have a two-channel Kondo effect at low energy, with the two channels being  the left-moving spin degrees of freedom
 in the upper and lower chains. The two equal Kondo coupling constants grow under the renormalization group and flow to the two-channel Kondo fixed point. It is a peculiarity of the two-channel Kondo effect that the two-channel fixed point corresponds to nothing but the exchange of the two channels~\cite{AffleckLudwig1991,MaldacenaLudwig1997,eggert1992}, here represented by the left-movers in the upper and the lower chain.
This situation is illustrated in the right panel of Fig.~\ref{SpinChains1}.

As the next step in our construction, we can imagine introducing additional impurity spins between different pairs of chains, as depicted in the left panel
of Fig.~\ref{SpinChains2}. We initially keep the horizontal spacing of adjacent impurities large. Owing to the chiral nature of the low-energy degrees of freedom
in the Kondo effect, backscattering is absent and the impurities do not affect each other. We can thus add them one-by-one, and eventually arrive in
the situation shown in right panel of Fig.~\ref{SpinChains2}. We can then let the distances shrink until we arrive at a close packing of the impurities,
as shown in Fig.~\ref{SpinChains3}. At this point, the weaving of the left-movers has led to two  stacks of lines  of left-movers
at an angle of $\pm 120^\circ$  relative
to the stack of of lines of right-movers (which remained unaffected by the coupling to the intersitial spins).
The chiral two-channel Kondo building block and the "weaving" construction therefore allow us to build up the staggered Kagome model from a stack of parallel antiferromagnetic spin chains.

An additional complication in intermediate steps of our construction arises from the fact that the Quantum Impurity problem for a single SU(2) spin-$1/2$ impurity coupled to two neighboring spin chains,
as shown for example in Fig.~\ref{SpinChains1}, contains a marginal operator. However, as discussed earlier, the model of interest (Eqn.~\eqnref{eqn:SpinH})
has the important property that the Hamiltonian changes sign under reflection of the horizontal direction.
This symmetry property is not respected by this marginal operator, which is thus forbidden in the model of
interest. At the level of our weaving construction, this symmetry property can be recovered at the final stage, shown in Fig.~\ref{SpinChains3}, by letting the coupling
on each Heisenberg chain, i.e. amongst the $\vec{S}$ as well as amongst the $\vec{T}$ spins, vanish while retaining only the three-spin couplings
(shown e.g. in Eq.(\ref{SpinForm2channel}))
 between the spins $\vec{S}$, the spins $\vec{T}$, and the impurity spins $\vec{S}_{imp}$.

It should be noted that this limit of taking the coupling along each chain to zero while retaining only the three-spin couplings is difficult to implement formally, since the approach of coupling the left- and right-moving low-energy modes depends on these modes being well-defined. We therefore implicitly rely on the three-spin coupling, when densely packed, to give rise to the same low-energy modes. In the case of the free fermion model, as explained in the next section and in particular in App.~\ref{app:MajoranaModel}, this limit can be treated exactly and we can show that the desired behavior emerges.

Furthermore, given that the arguments above are not rigorous, one may wonder if competing orders can emerge that break translational or other symmetries of the system. While a full analytical treatment of symmetry-breaking terms goes beyond the scope of this manuscript, we refer the reader to Refs.~\onlinecite{Messio2011,Pereira2018} for closely related discussions. Furthermore, our numerical results discussed in Sec.~\ref{sec:quasi1d} indicate that no symmetries of the system are spontaneously broken.

%%%%%%%%%%%%%%%%%%%%%%%%%%%%%%%%%%%%%%%%%%%%%%%%%%%%%%%%%

\subsection{Validation for the Majorana model}

\label{SubSectionValidationForTheMajoranaModel}

We can verify the validity of the technical arguments of the weaving construction for the spin model, laid out in the previous section, by again considering the Majorana model \eqref{eqn:MajH}. Here we can establish {\it rigorous} results for every single step of the weaving construction and demonstrate that it indeed builds up the correct two-dimensional gapless state with its distinct Fermi lines as shown in Fig.~\ref{fig:bands}.

The starting point of the weaving construction for the Majorana model is again a limit where we consider decoupled chains of Majorana zero modes. Collectively, such a one-dimensional system of interacting Majorana zero modes forms a gapless system described by a Ising conformal field theory \cite{KitaevChain,GruzbergReadLudwig,GoldenChain} consisting of left and right-moving Ising degrees of freedom
(in full analogy to the gapless Heisenberg chain of the spin model). 
Coupling these gapless left and right movers to  interstitial Majorana zero modes via the ``bowtie'' interactions of the type
 shown in Figs.~\ref{SpinChains1},
again turns out to leave the right movers uncoupled (modulo RG-irrelevant terms), while the left movers are interwoven in the same way as we have discussed for the spin model. This is
 discussed in detail in Appendix~\ref{app:MajoranaModel}.
 With this most crucial step of the weaving construction in place, we note that 
 chirality again prohibits backscattering and we can therefore proceed along the same line of argument as for the spin model in adding further interstitial Majorana zero modes one-by-one to ultimately build up the entire Kagome lattice (again, as shown in Figs.~\ref{SpinChains2} and \ref{SpinChains3}).
This again leads us to the three stacks of parallel lines of gapless {\it chiral}  modes illustrated in the right panel of Fig.~\ref{SpinChains3}. For the Majorana model, the weaving construction is thus shown to rigorously yield the Fermi surface (Fig.~\ref{fig:bands}) that one also obtains by direct diagonalization of the Kagome system in the thermodynamic limit.

This success of the weaving construction in reproducing the Fermi surface of the fully coupled Majorana model on the Kagome lattice lends further support to its 
validity  in the context of the original spin model.

%%%%%%%%%%%%%%%%%%%%%%%%%%%%%%%%%%%%%%%%%%%%%%%%%%%%%%%%%

\subsection{Connection to coupled wire construction and sliding Luttinger liquids}

We close this Section by noting that our weaving constructions bears some resemblance to the coupled wire constructions of Kane and collaborators \cite{Kane2002,KaneTeo2014}. But while the coupled wire construction
is typically used to describe a gapped state (with chiral edge modes) such as the closely related Kagome model \cite{Bauer2014} harboring a chiral spin liquid \cite{Meng2015,Gorohovsky2015}, our construction explicitly aims at describing a two-dimensional, gapless bulk state.

On a more technical level, both approaches start from a set of decoupled gapless chains and proceed by coupling adjacent chains. In the coupled wire construction a pair of right and left movers of adjacent chains are gapped, with   gapless modes remaining only at the system boundary.
In contrast,  in our construction the notion of adjacency, i.e. which pairs of gapless modes are de facto coupled, changes as interstitial impurities are added one-by-one, see e.g. Fig.~\ref{SpinChains3}, and no modes get ever gapped out (but get instead redirected). 

Another seeming similarity of our model, and the approach described here, is to sliding Luttinger liquid phases~\cite{Mukhopadhyay2001,Mukhopadhyay2001b,Vishwanath2001}.
However, there are important conceptual differences to our work. First, while the weaving construction described above is anisotropic, the model as well as the low-energy gapless phase
that are recovered at the end are isotropic, i.e. fully respect $120^\circ$ rotations of the lattice.
Furthermore, the low-energy modes on each stack of parallel chains are chiral. This is crucial for the stability of the phase, since it prevents backscattering
on a single chain that would gap out left- and right-moving modes.

%%%%%%%%%%%%%%%%%%%%%%%%%%%%%%%%%%%%%%%%%%%%%%%%%%%%%%%%%

\section{Numerical analysis of ladder systems}
\label{sec:quasi1d}

We now turn our attention to quasi-one-dimensional realizations of the interacting spin model of Eqn.~\eqnref{eqn:SpinH}.
These quasi-one-dimensional systems are more amenable to simulations using matrix-product
states~\cite{fannes1992,white1992,white1992-1,ostlund1995,schollwoeck2005,schollwoeck2011,stoudenmire2012}, which allow us to make
reliable numerical statements about the spectrum of these systems as well as, in the case where they are gapless, the nature of the associated
conformal field theory. As outlined already in the case of free fermions in Sec.~\ref{sec:ff1d},
we expect the number of gapless modes in these quasi-one-dimensional realizations to scale linearly with the width of the system.

\begin{figure}[b]
  \includegraphics[width=2.5in]{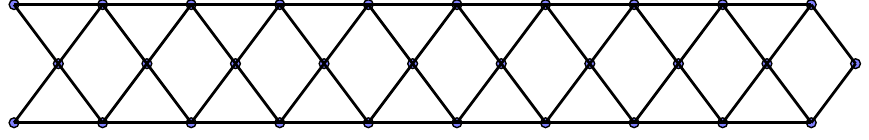}
  \includegraphics[width=2.5in]{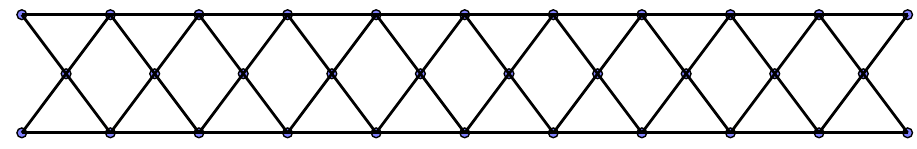}
  \caption{{\bf Boundary conditions} used in our numerical analysis of narrow cylinders.
  Top: periodic boundary conditions (in the long direction).
  Bottom: open boundary conditions. \label{fig:clusters} }
\end{figure}

To check for the stability of the gapless modes, we consider a perturbed version of the Hamiltonian, given by
\begin{equation} \label{eqn:SpinH-perturbed}
H_K = K H + (1-K) \sum_{\langle i,j \rangle} \vec{S}_i \cdot \vec{S}_j
\end{equation}
on the lattice as shown in Fig.~\ref{fig:clusters}. Here, the first term is the original
 spin Hamiltonian of Eqn.~\eqnref{eqn:SpinH}, and the sum in the second terms runs only over bonds
shown as horizontal lines in Fig.~\ref{fig:clusters}, i.e. along the chains. In the limit $K = 0$, the system consists
of two uncoupled Heisenberg chains and a number of completely decoupled spins; for $K=1$, only the chiral
term is present, as in
Eq.(\ref{eqn:SpinH}).
While one could in principle include the Heisenberg couplings on all bonds and not just horizontal bonds, this leads in the limit of $K=0$ to a model
that was recently  shown to harbor a very exotic gapless phase~\cite{Amir2017}, which would make the analysis more involved.

\subsection{Ground states of the two-leg cylinder}
\label{sec:quasi1dw2}

We first consider a cylinder of circumference 2, as discussed also in Sec.~\ref{sec:ff1d}. The clusters we study are shown in Fig.~\ref{fig:clusters};
the number of sites is $N=3n$ for periodic systems and $N=3n+2$ for open boundary conditions, where we choose $n$ even
to have an even number of sites. We perform DMRG~\cite{fannes1992,white1992,white1992-1,ostlund1995,schollwoeck2005,schollwoeck2011,stoudenmire2012}
simulations using the ALPS~\cite{bauer2011-alps} MPS code~\cite{dolfi2014}. While the model exhibits an SU(2) spin symmetry, our
simulations only exploit a U(1) subgroup corresponding to the total magnetization, $S_z$.

\paragraph{Energy spectrum}

\begin{figure}
  \includegraphics{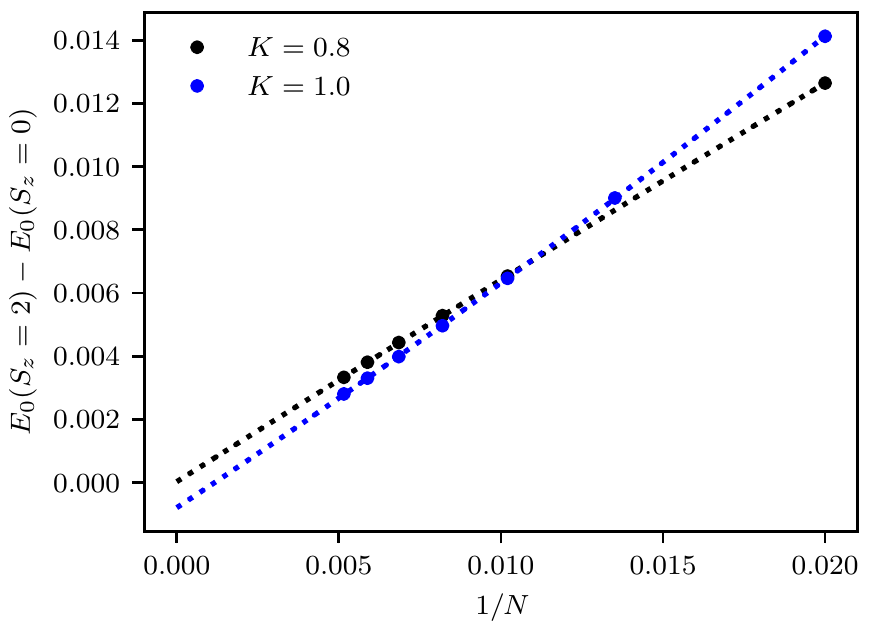}
  \includegraphics{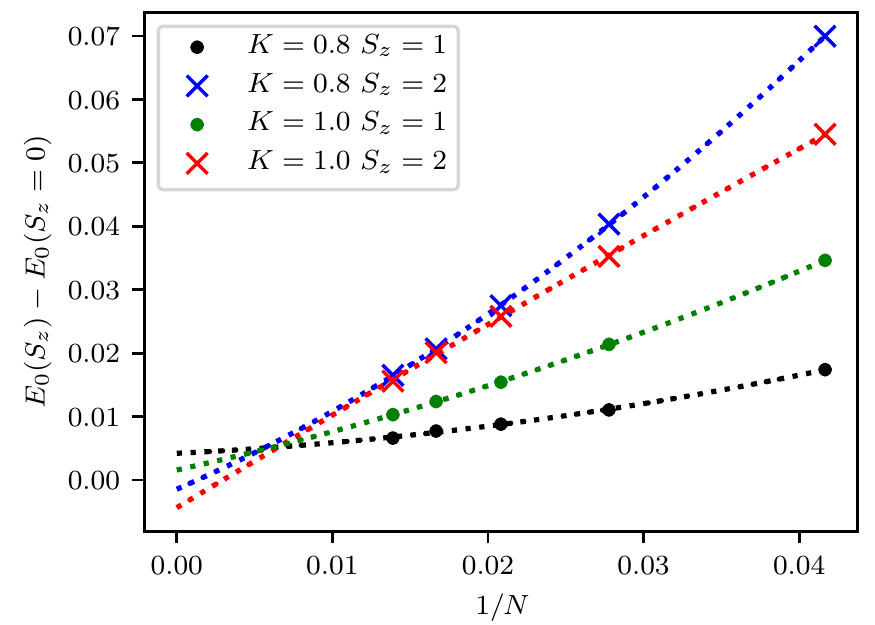}
  \caption{{\bf Spectral gap} of the $W=2$ cylinder with open (top panel) and periodic (bottom panel) boundary conditions.
  Fits to $E=E_0 + a/L + b/L^2$ are indicated by the dotted lines.
  \label{fig:w2_spectrum} }
\end{figure}

In Fig.~\ref{fig:w2_spectrum}, we analyze the excitation spectrum of \eqnref{eqn:SpinH-perturbed} both in the ideal limit $K=1$
and in a perturbed case $K=0.8$, i.e. with a Heisenberg term of strength $J=1-K=K/4$. The notation used here is that
$E_n(S_z)$ is the energy of the $n$'th-lowest state in a fixed $S_z$ sector. We find that for open boundary conditions,
the lowest energy in the $S_z=1$ sector is degenerate with the ground state in the $S_z=0$ sector, indicating either that
there is a degenerate SU(2) triplet, or that the ground state itself is an SU(2) triplet and not a singlet.
A similar degeneracy is also observed in the free-fermion variant of the model. However, since in a critical system all excitation
energies are expected to collapse to 0, we can equally well analyze the quintuplet ($S=2$) gap, which is given by $\Delta_2 = E_0(S_z=2)-E_0(S_z=0)$
and shown in the top panel of Fig.~\ref{fig:w2_spectrum}. We find that
the quintuplet gap collapses to 0 as the system size is increased in very good agreement with $\Delta_2 \sim a/N + b/N^2$, where the quadratic coefficient
$b$ is very small. This holds for both values of $K$, and is strong indication that the system is gapless with dynamical critical exponent
$z=1$ for both $K=1$ and $K=0.8$.

\begin{figure}[t]
  \includegraphics{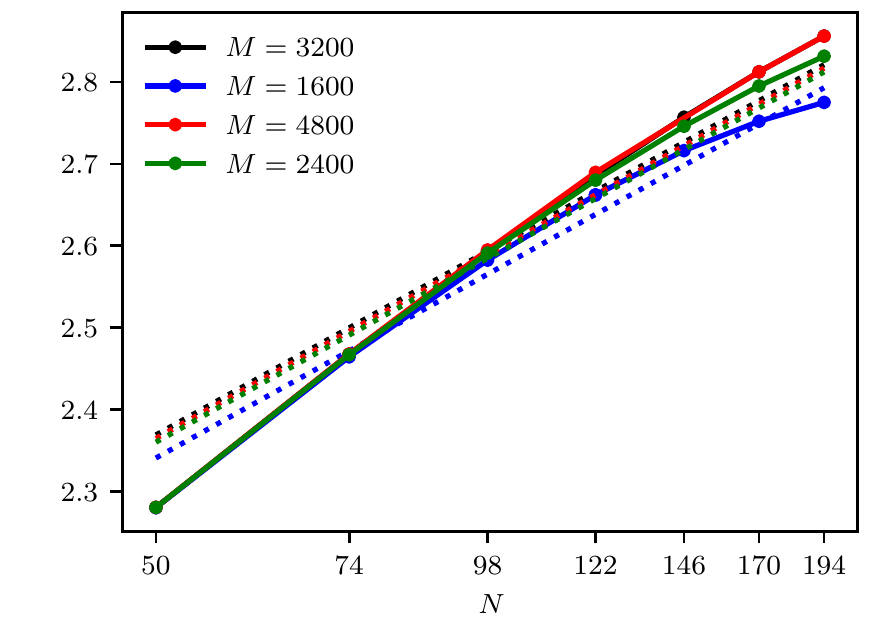}
  \includegraphics{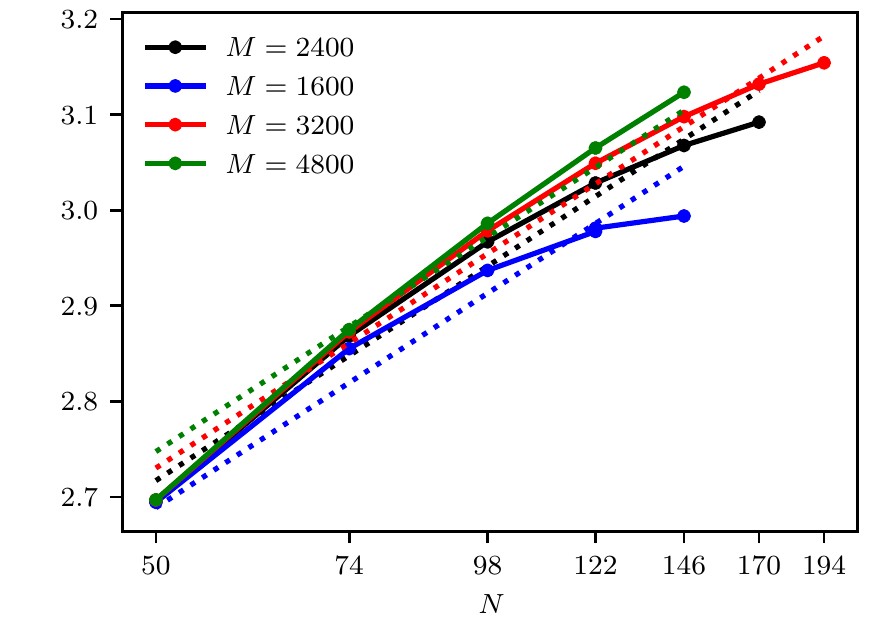}
  \caption{{\bf Entanglement entropy} of the $W=2$ cylinder with open boundary conditions, with $K=1.0$ (top panel)
  and $K=0.8$ (bottom panel). In this parametrization, the strength of the Heisenberg term is $J=1-K$.
  In both panels, the dotted line is a fit to $S(N)=(2/6)\log (N/2) + S_0$.
  \label{fig:w2cylinder_open_entropy} }
\end{figure}

To further substantiate this, we analyze in the bottom panel of Fig.~\ref{fig:w2_spectrum} the fully periodic case, i.e. long tori. In
this case, there is no degeneracy between the $S_z=0$ and $S_z=1$ sectors (i.e. the ground state is a singlet and there is a finite
triplet gap), and we can thus analyze both the triplet and quintuplet gaps. In all cases, we again find good agreement with
$a/N+b/N^2$. While in this case the quadratic term is more pronounced, this is likely due to microscopic details and does not invalidate
the conclusion of a gapless system with $z=1$.

\paragraph{Entanglement entropy}

A more refined prediction is that for the cylinder of width two, the system should exhibit two gapless modes. To confirm this, we measure
the central charge of the conformal field theory that forms the long-wavelength description of a critical system with $z=1$ 
(confirmed in the previous subsection Sect. \ref{sec:quasi1dw2} a.)
using the
well-known relationship between the logarithmic divergence of the entanglement entropy and the central charge~\cite{vidal2003,calabrese2004}.
This approach is based on the scaling of the bipartite entanglement entropy, which for blocks of size $n$ in a system of total size $N$, where $n \ll N$, is known to asymptote to
\begin{equation}
S(n) = \frac{c}{6} \log n + S_0.
\end{equation}
Here, $S_0$ is a non-universal constant and $c$ is the desired central charge. When the condition $n \ll N$ is not satisfied, additional
subleading and possibly oscillating corrections appear, see e.g. Ref.~\onlinecite{laflorencie2006}. In practical simulations, the central
charge can be obtained either by fixing $N$ and varying $n$, or by fixing $n$ (say, to $n=N/2$) and varying $N$. In the systems at hand,
since $S(n)$ for fixed $N$ exhibits very strong and complicated oscillatory terms (not just even-odd effects), we focus on the second
approach with $n=N/2$.

Our numerical results are shown in Fig.~\ref{fig:w2cylinder_open_entropy}.
Since the systems we study are highly entangled, the measured entanglement entropy exhibits a strong dependence on the bond dimension
even for very large bond dimensions of several thousand. We therefore plot results for various bond dimensions, but fit each curve
to the same shape $S = (2/6) \log (N/2) + S_0$, where $S_0$ is the only fit parameter. It can be seen that for both $K=1$ and $K=0.8$, the
curves approach this scaling as the system size and the bond dimension are increased. The remaining curvature visible on the
logarithmic scale is due to a combination of the limited bond dimension and subleading contributions to the entropy scaling.
In summary, our results confirm the presence of two gapless modes. (The Lieb-Schultz-Mattis-Hastings Theorem would only predict
the presence of a gapless mode, which would generically have $c=1$ in our system.)

\subsection{Gutzwiller-projected wavefunctions}
\label{sec:Gutzwiller}

For cylinders of circumference larger than $W=2$, we find that DMRG does not reliably converge to the ground state. We instead
pursue an approach based on a variational wavefunction obtained by Gutzwiller-projecting a free-fermion Slater determinant for
4 species of Majorana fermions. To this end, we exploit the close analogy to the Majorana system described by the Hamiltonian~\eqnref{eqn:MajH}.
While there is no rigorous microscopic connection between the Majorana system and the system of spins that is our main interest, the
shared symmetries and closely related physical behavior of lines of gapless excitations is suggestive that the Majorana Hamiltonian
may be a useful starting point for constructing ansatz states that correctly capture the universal behavior of the spin system. Our goal
here is thus to construct these spin states on lattices where exact simulations of the ground state are prohibitive, and then numerically
confirm that the universal physics -- which in this case is also captured by the number of gapless modes in a quasi-one-dimensional
geometry -- is correctly captured in these ansatz states.
We note that this approach is related to the one of Ref.~\onlinecite{biswas2011}, where a parton construction of a related model
on the triangular lattice as well as the low-energy properties of the spin liquid phase are discussed. 

We consider a system comprised of four identical copies of the Majorana Hamiltonian
of Eq. (\ref{eqn:MajH}), and thus described by
\begin{align} \label{eqn:ManyMaj}
	H = \sum_{\alpha=1}^4 \left[ \sum_{i,j,k \in \bigtriangleup} \tilde{\chi}^\alpha_{ijk} - \sum_{i,j,k \in \bigtriangledown} \tilde{\chi}^\alpha_{ijk} \right],
\end{align}
where $\tilde{\chi}^\alpha_{ijk} = \i (\gamma_i^\alpha \gamma_j^\alpha + \gamma_j^\alpha \gamma_k^\alpha + \gamma_k^\alpha \gamma_i^\alpha)$.
From the four Majorana fermions on each site, we can form a complex spinful fermion $c_{i \sigma}$ on the $i$'th site as
\begin{align}
c_{i \uparrow} &= \gamma_i^1+i \gamma_i^2 &c^\dagger_{i \uparrow} &= \gamma_i^1-i \gamma_i^2 \\
c_{i \downarrow} &= \gamma_i^3+i \gamma_i^4 &c^\dagger_{i \downarrow} &= \gamma_i^3-i \gamma_i^4.
\end{align}
To obtain a variational spin wavefunction, we apply a Gutzwiller projection to the ground state of this system using the standard projection operator
(using $n_{i \uparrow}=c_{i \uparrow}^\dagger c_{i \uparrow}$, $n_{i \downarrow}= c_{i \downarrow}^\dagger c_{i \downarrow}$)
\begin{equation} \label{eqn:gutzproj}
\mathcal{P} = \prod_i \left( n_{i \uparrow} - n_{i \downarrow} \right)^2
\end{equation}
which selects states with exactly one fermion per site. In the language of the original fermions, this corresponds to
\begin{equation}
\mathcal{P} = \prod_i \frac{\gamma_i^1 \gamma_i^2 \gamma_i^3 \gamma_i^4 -1}{2},
\end{equation}
i.e. the projection into the odd-occupancy space on each site. The SO(4) symmetry
of the original problem, which is manifest as SU(2)$\times$SU(2) symmetry on the level of the spinful fermions, is reduced by
the projection, and only the spin SU(2) symmetry remains.

To carry out our calculations, we again work with a formalism of matrix-product states (MPS). We first diagonalize the Hamiltonian~\eqnref{eqn:ManyMaj}
and compute the correlation matrix $C_{ij} = \langle c_i^\dagger c_j \rangle$ of its ground state. We then use a recently proposed method~\cite{fishman2015}
to obtain an MPS representation of this ground state. Once we have obtained this MPS, the Gutziller projection as well as measurement of the entropy
can be performed using standard techniques; see App.~\ref{app:numerics} for details.

\begin{figure}
  \includegraphics{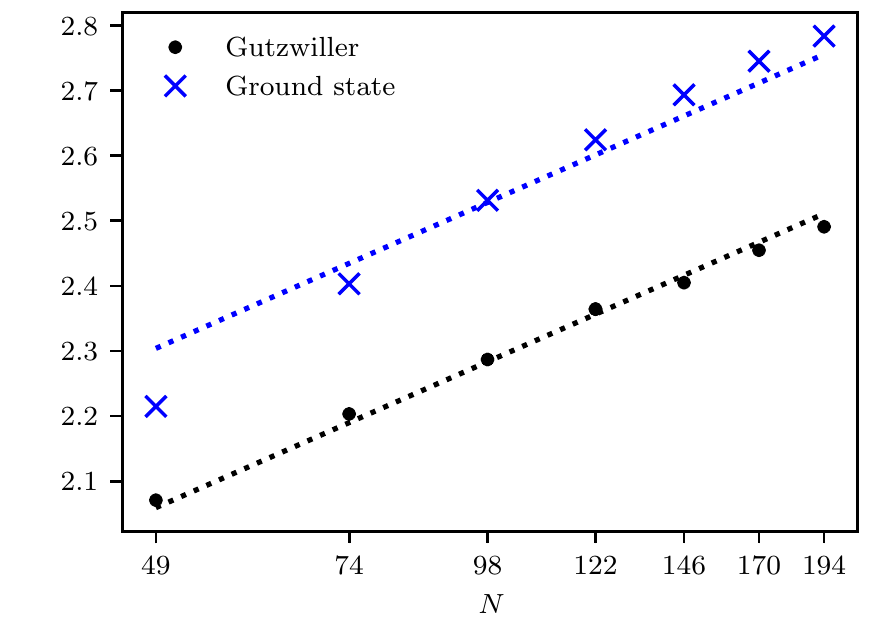}
  \includegraphics{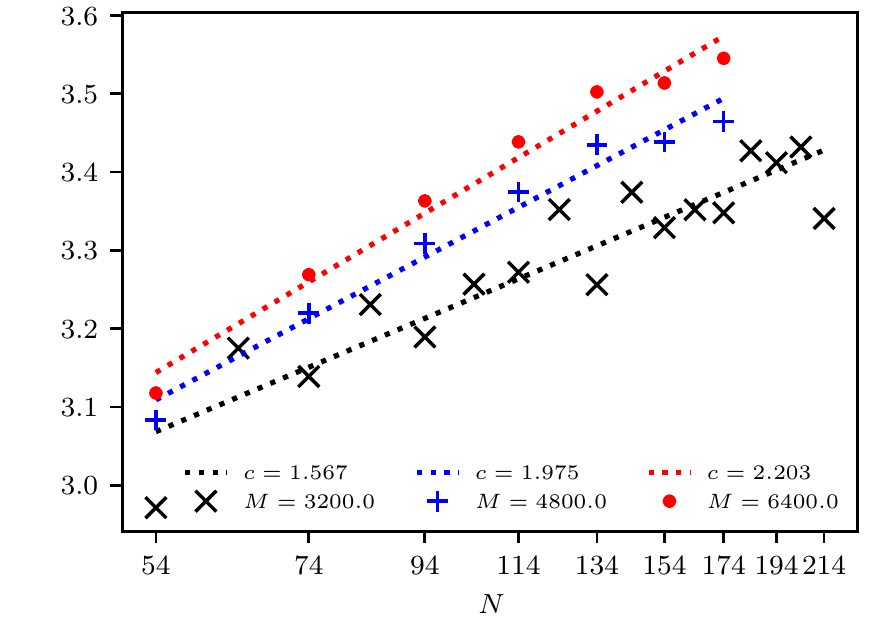}
  \caption{{\bf Entanglement entropy} for middle cuts of the Gutzwiller-projected ground states of~\eqnref{eqn:ManyMaj}. The top panel shows
  results for the same $W=2$ cylinder as discussed in Sec.~\ref{sec:quasi1d}, while the bottom panel shows results for a strip of
  width $W=3$. In the top panel, the dashed lines are best fits to central charge $c=2$; in the bottom panel, the central charge is a
  free parameter of the fit and the legend indicates the optimal value for each bond dimension $M$ ($M=2400$ in the top panel).
  \label{fig:gutzwiller} }
\end{figure}

Our results are summarized in Fig.~\ref{fig:gutzwiller}. We first consider the same $W=2$ cylinder as in the previous section, which was there argued
to form a gapless system with two gapless modes, and hence central charge $c=2$. These results, for bond dimension $M=2400$, are shown
again in the top panel of Fig.~\ref{fig:gutzwiller}, where they are compared to the entropy of the Gutzwiller-projected state in the exact same
geometry and using the same bond dimension. We see that while the entropy for each system size is systematically larger in the ground state,
the scaling with system size is remarkably similar, and both exhibit good agreement with $S \sim \frac{2}{6} \log N$. The shift in the entropy
can likely be explained with non-universal short-range physics that is not accurately captured in the Gutzwiller-projected ansatz state, while
being fully captured in the numerically optimized ground state.
While it is in principle possible to introduce free parameters into the Hamiltonian~\eqnref{eqn:ManyMaj} and then optimize these to obtain
ansatz states with lower ground state energy, we do not attempt this numerically challenging procedure here.

Having built trust in our ability to capture the universal long-distance physics of our spin system in the framework of Gutzwiller-projected
wavefunctions, we now move on to a strip (fully open system) of width $W=3$. For this system, we expect three gapless modes, i.e.
a central charge of $c=3$. We use open boundary conditions in both directions because there are no natural Kagome cylinders with
odd circumference. Performing DMRG simulations of this system has yielded converged ground states only for very small systems
and thus has not allowed us to extract the scaling of the entanglement entropy with system size.

Our results for Gutzwiller-projected wavefunctions can be seen in the lower panel of Fig.~\ref{fig:gutzwiller} for three different
values of the bond dimension ranging from $M=3200$ to $M=6400$. We also show a fit to the usual form
$S = S_0 + \frac{c}{6} \log N$. We find that the entropy, and thus the estimate for $c$, grows strongly with the bond dimension.
Our estimates should thus be seen as lower bounds on the physical values. For the largest bond dimension accessible to our
simulations, $M=6400$, we find an approximate central charge of $c=2.2$, and thus significantly greater than 2. This is strongly
suggestive that the physical central charge of the system is $c > 2$, as theoretically predicted, and provides support that the
universal behavior of the phase described above extends to the two-dimensional limit of the spin system.

%%%%%%%%%%%%%%%%%%%%%%%%%%%%%%%%%%%%%%%%%%%%%%%%%%%%%%%%%%%%%%%%%%

\section{Three-dimensional extensions}
\label{sec:3d}

\begin{figure}[t]
  \includegraphics[width=0.8\columnwidth]{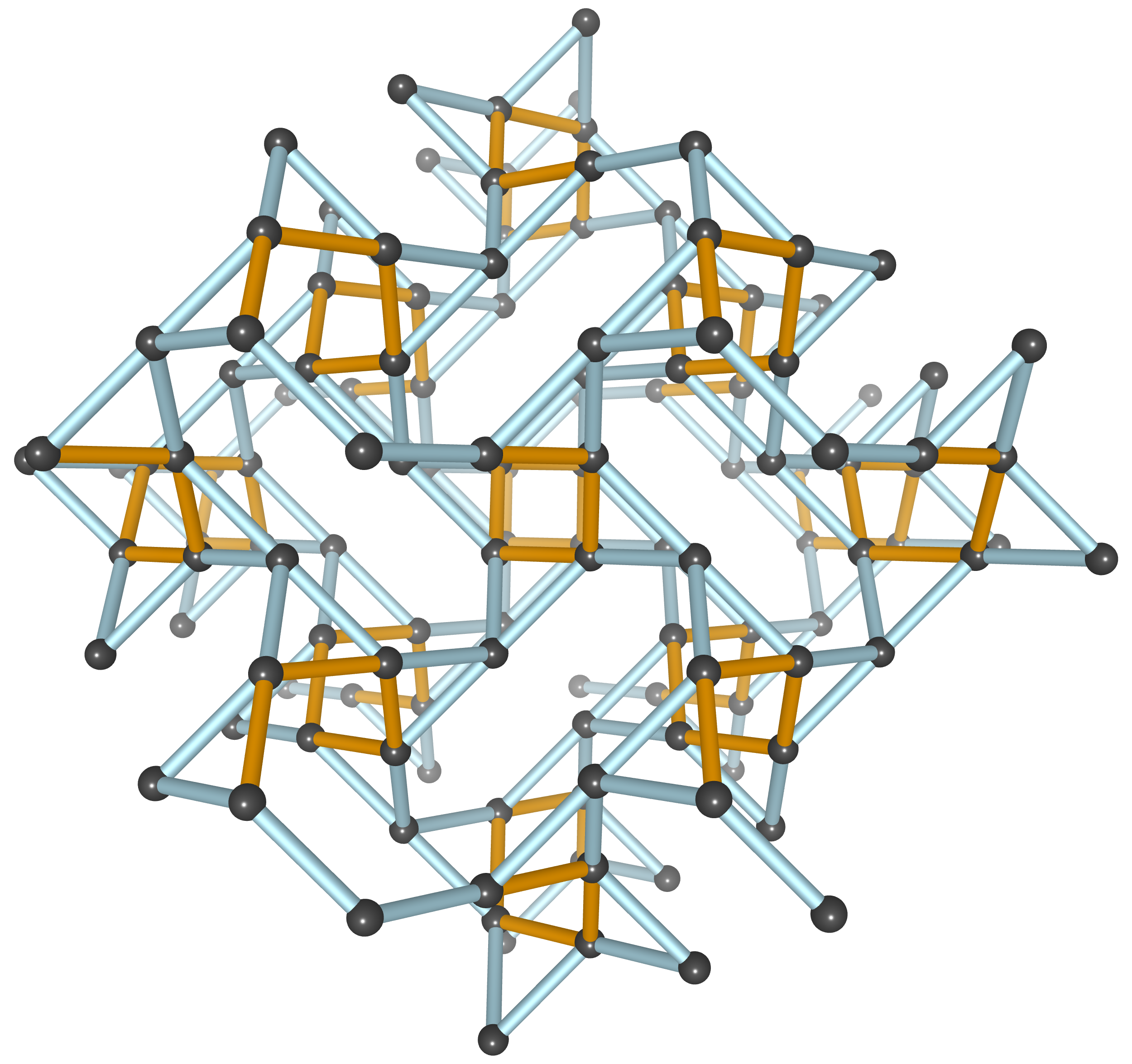}
  \caption{The {\bf hyperkagome lattice} of corner-sharing triangles. 
  	        Its chiral cubic lattice structure consists of two types of counterrotating helices, 
	        a ``square" helix of four triangles performing one screw rotation (indicated by the brown links) 
	        and an ``octagon" helix of eight triangles performing a counterrotating screw rotation (indicated by the blue links).
	        The medial lattice of the hyperkagome lattice is a tricoordinated lattice structure, which is referred to 
	        as Laves graph, K$_4$ crystal or (10,3)a lattice. Further details of these lattices are provided in 
	        Appendix \ref{app:lattices}.
  \label{fig:hyperkagome} }
\end{figure}

Our discussion of the physics underlying the formation of gapless spin liquids on the Kagome lattice has at no point relied on the fact
that the Kagome lattice is in fact a {\em two-dimensional} lattice. So it is a natural question to ask whether we can extend these ideas to spin models that are defined for three-dimensional (3D) generalizations of the Kagome lattice. Probably the best known example of such a 3D generalization is the hyperkagome lattice, which has been introduced in the context of the iridate spin liquid material Na$_4$Ir$_3$O$_8$ \cite{Okamoto2007}. Like the Kagome lattice the hyperkagome lattice is a lattice of corner-sharing triangles as illustrated in Fig.~\ref{fig:hyperkagome}. It can be conceptualized as a cubic lattice structure of two types of counterrotating helices,
a ``square" helix of four triangles performing one screw rotation (indicated by the brown links in the figure) and an ``octagon" helix of eight triangles performing a counterrotating screw rotation (indicated by the blue links in the figure). This chiral  structure becomes even more evident when going to the medial lattice, i.e. by contracting every triangle to a single 
tricoordinated lattice site. 
This medial lattice of the hyperkagome lattice is well known in the literature and referred to
as Laves graph \cite{Laves1993}, K$_4$ crystal \cite{K4}, or hyperoctagon lattice \cite{Hermanns2014} and part of 
a larger family of three-dimensional, tricoordinated lattices that have been systematically classified by A.~F.~Wells \cite{Wells1977}. Following Well's classification via Schl\"afli symbols, the hyperkagome lattice is the premedial lattice of lattice (10,3)a, indicating that all elementary closed loops of triangles in the hyperkagome lattice are of length 10. This also points a route towards further 3D  generalizations of the Kagome lattice, which can be constructed as premedial lattices of 
Wells' tricoordinated 3D structures
\footnote{The fermionic tight-binding band structures associated with these lattices have been extensively classified and
demonstrated to exhibit nodal structures of points, lines, and surfaces depending on the underlying lattice structure \cite{OBrien2016,Wawrzik2018}}. 
Two more such generalizations of corner-sharing triangles for medial lattices (10,3)b and (10,3)c are given in Appendix \ref{app:lattices}, including a side-by-side visualization and details of their respective unit cells.

Here we concentrate on the hyperkagome lattice. We can readily generalize the spin model of Eq.~\eqref{eqn:SpinH} by first observing that we can indeed assign a fixed chirality to all triangles of the lattice. As discussed in detail in Appendix \ref{app:hyperkagome} there are eight independent triangles per unit cell (of 12 sites), which in principle allows to define some distinct $2^{8-1}$ possible assignments of spin chirality terms to these triangles (up to an overall sign).
Out of this multitude of assignments, there is one distinct choice for a model where the Hamiltonian
anticommutes with the 180 degree rotational symmetries of the lattice. Visually, this particular choice corresponds
to the situation where whenever two triangles meet and two of their edges are parallel the assigned direction of these 
edges coincides, akin to the assignment illustrated for the staggered Kagome model in Fig.~\ref{SpinChains3}. 

\begin{figure}
  \includegraphics[clip, trim= 1.5cm 1.5cm 1.5cm 1.5cm,width=\columnwidth]{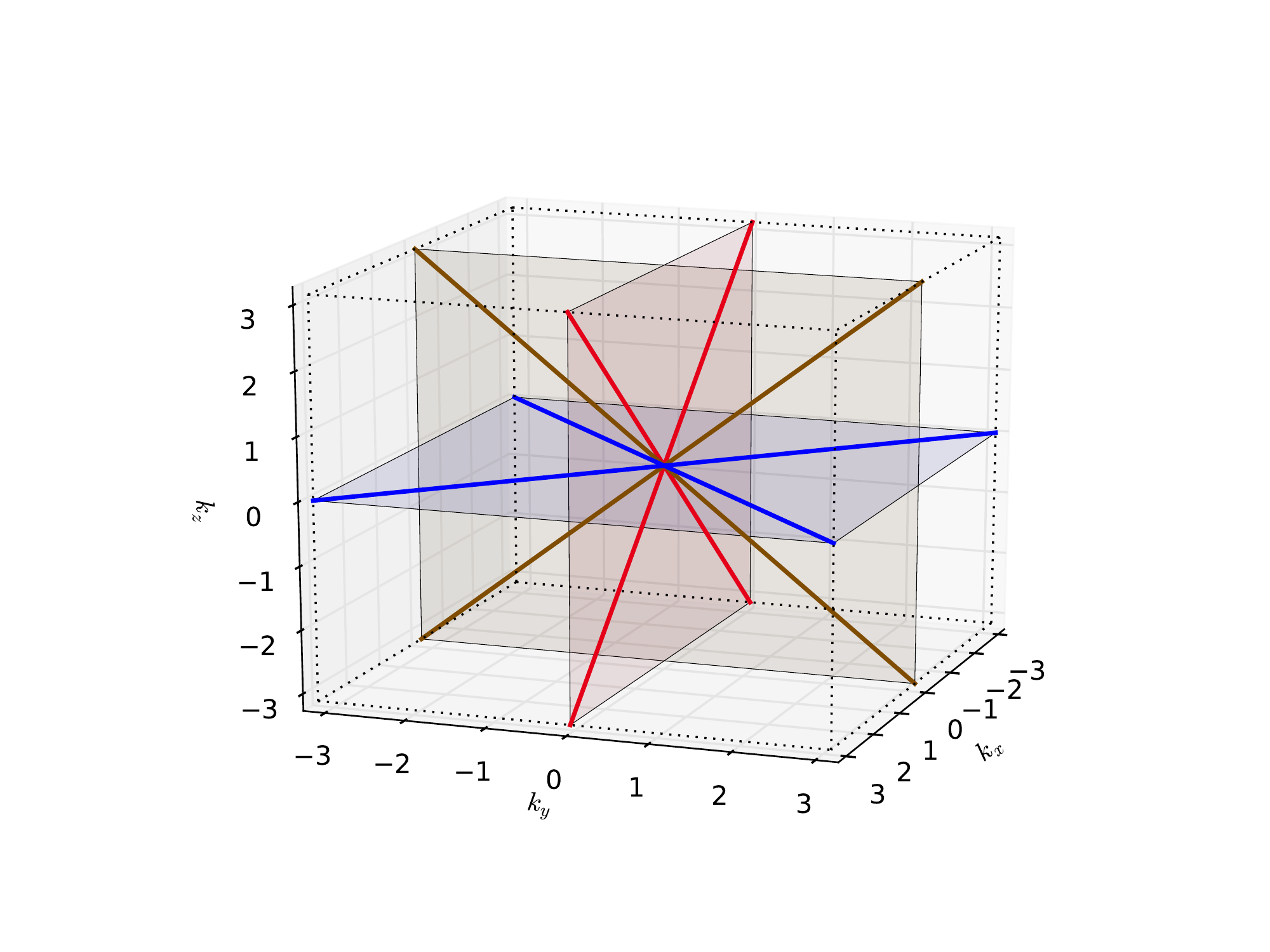}
  \caption{Symmetry-protected \textbf{gapless lines} on the hyperkagome lattice. \label{fig:3Dlines} }
\end{figure}

While we cannot solve the 3D spin model directly, we can follow a similar route as in two spatial dimensions. 
First, we note that the particular choice of chirality assignments and resulting edge directions in principle allows for a chiral 
Kondo lattice construction as discussed in Sec.~\ref{sec:weaving}.
Second, we can educate ourselves about the expected physics by considering the corresponding Majorana fermion model of Eq.~\eqref{eqn:MajH} in this three-dimensional setting. Here we note that we can again apply the symmetry arguments of Sec.~\ref{sec:symmetry} to find that the 180 degree rotational symmetries protect a total of six gapless lines in momentum space as illustrated in Fig.~\ref{fig:3Dlines}. For details of this calculation, we refer the reader to Appendix \ref{app:hyperkagome}.

%%%%%%%%%%%%%%%%%%%%%%%%%%%%%%%%%%%%%%%%%%%%%%%%%%%%%%%%%%%%%%%%%%

\section{Summary}
\label{sec:summary}

To summarize,  this manuscript has laid out a detailed proposal for the description of non-Fermi liquid physics in the context of gapless quantum spin liquids possibly arising in Kagome quantum magnets in two and three spatial dimensions. These quantum spin liquids of interest here exhibit gapless spinon degrees of freedom not only at singular points in momentum space (such as a Dirac spin liquid), but along {\em lines} in momentum space. As such, these spin liquid systems exhibit a number of characteristics akin to a metal, e.g.  non-vanishing thermal transport down to zero temperature and a specific heat in temperature that directly reflects the gapless manifold of spinons, and are therefore also referred to as Bose metals.

Our description of these spinon surfaces originates from a formulation of the original Kagome spin model in terms of  a chiral $2$-channel Kondo lattice model. In particular, we have developed a novel ``weaving construction" for gapless bulk states, somewhat similar to the coupled wire construction for gapped quantum states. We supported the validity of this approach by demonstrating the explicit reconstruction of the 2D Fermi surface of a closely related Majorana fermion model in terms of a chiral resonant level (Kondo) lattice model, which is fully analytically tractable.
 
More generally, we have identified a possible symmetry protection mechanism of these gapless spinon surfaces, which can also be extended to three spatial dimensions in a straightforward manner. This allowed us to identify possible 3D Kagome-like lattice models that might harbor gapless spin liquids with a two-dimensional spinon surface.

In a recent manuscript~\cite{Pereira2018}, Pereira and Bieri discuss a related model for a gapless Kagome spin liquid, which has been inspired by our original, now superseded preprint~\cite{bauer2013}.
Using a coupled-chains approach,
rather different from the methods employed in the present article,
 they argue for the presence of stable gapless lines in their model, similar to what we discuss here.

%%%%%%%%%%%%%%%%%%%%%%%%%%%%%%%%%%%%%%%%%%%%%%%%%%%%%%%%%%%%%%%%%%

\acknowledgments
S.T. acknowledges discussions with Kevin O'Brien and Maria Hermanns on three-dimensional generalizations of the Kagome lattice.
This work is supported in part by the NSF under grant No. DMR-1309667 (A.W.W.L.).
S.T. acknowledges partial support by the DFG within the CRC network TR 183 (project B01)
and the hospitality of Microsoft Station Q during the final stages of this work.
We thank the Aspen Center for Physics for hospitality.

\appendix

%%%%%%%%%%%%%%%%%%%%%%%%%%%%%%%%%%%%%%%%%%%%%%%%%%%%%%%%%

\section{Chiral Kondo coupling}
\label{app:ChiralKondo}

Consider a one-dimensional antiferromagnetic Heisenberg spin chain. A single spin operator $\vec{S}_n$ at site $n$ of a gapless spin $1/2$ antiferromagnetic chain may be expressed in terms of the long-wavelength degrees of freedom of the $SU(2)_1$ WZW conformal field theory, which vary slowly on the lattice scale, as 
\be
a^{-1} \vec{S}_n \rightarrow \vec{J}_L(z_n) + \vec{J}_R(\bar{z}_n) + i(-1)^n\mbox{const} \times \mbox{tr}(g(z_n,\bar{z}_n)\vec{\sigma})
\ee
where $g$ is an $SU(2)$ matrix that describes the continuum spin degrees of freedom, $\vec{J}_{L,R}$ are Kac-Moody currents of the $SU(2)_1$ WZW conformal field theory, and $a$ is the lattice spacing. The constant "$\mbox{const}$" is non-universal. The holomorphic and anti-holomorphic coordinates $z_n$ and $\bar{z}_n$ correspond to the space and imaginary time coordinates $z_n=\tau+ix_n$ and $\bar{z}_n=\tau-ix_n$ where $x_n=na$ (we have chosen units in which the velocity parameter is set to unity). Therefore, the interaction $\epsilon^{abc}S^b_n S^c_{n+1}$ may be expressed as follows in terms of the long-wavelength degrees of freedom.
\begin{eqnarray}
\nonumber 
&\epsilon^{abc}& [J_L^b(z_{n}) J_L^c(z_{n+1})  \\ \nonumber
&+&J_R^b(\bar{z}_{n}) J_L^c(z_{n+1}) + J_L^b(z_{n}) J_R^c(\bar{z}_{n+1}) + J_R^b(\bar{z}_{n}) J_R^c(\bar{z}_{n+1}) \\
\nonumber &+& i (-1)^n \mbox{const} \hspace{0.1cm} \mbox{tr}(g(z_{n},\bar{z}_{n})\sigma^b)(J_L^c(z_{n+1})+J_R^c(\bar{z}_{n+1})) \\
\nonumber &-& i (-1)^n \mbox{const} (J_L^b(z_{n})+J_R^b(\bar{z}_{n}))\mbox{tr}(g(z_{n+1},\bar{z}_{n+1})\sigma^c) \\
\label{3spin} &+& \mbox{const}^2\mbox{tr}(g(z_{n},\bar{z}_{n})\sigma^b)\mbox{tr}(g(z_{n+1},\bar{z}_{n+1})\sigma^c)]
\end{eqnarray}

The $J_LJ_R$ terms in the first line of Eq.~\eqref{3spin} may be simplified as follows:
\begin{eqnarray}
\nonumber && J_R^b(\bar{z}_{n}) J_L^c(z_{n+1}) + J_L^b(z_{n}) J_R^c(\bar{z}_{n+1}) \\
\nonumber &=&J_R^b(\bar{z}_{n}) (J_L^c(z_{n})+\sum_{m=1}^\infty\frac{(ia)^m}{m!}\partial^m J_L^c(z_{n})) \\ \nonumber
&+&J_L^b(z_{n}) (J_R^c(\bar{z}_{n})+\sum_{m=1}^\infty\frac{(-ia)^m}{m!}\bar{\partial}^m J_R^c(\bar{z}_{n}))  \\
\nonumber &=& J_R^b(\bar{z}_{n}) J_L^c(z_{n}) + J_L^b(z_{n}) J_R^c(\bar{z}_{n}) +O(a)
\end{eqnarray}

Since this term is symmetric in indices $b$ and $c$ it will sum to zero when contracted with the antisymmetric Levi-Civita symbol. The remaining $J_LJ_L$ and $J_RJ_R$ terms in the first line of Eq.~\eqref{3spin} may be simplified through their operator product expansion. For small $| z - \zeta |$ the Kac-Moody currents of $SU(2)_1$ obey the OPE
\begin{eqnarray}
\nonumber J^b_L(z) J^c_L(\zeta) &=& \frac{1}{2}\frac{\delta^{bc}}{(z-\zeta)^2}+\frac{i\epsilon^{bcd}}{z-\zeta} J^c_L(\zeta) \\
\nonumber J^b_R(\bar{z}) J^c_R(\bar{\zeta}) &=& \frac{1}{2}\frac{\delta^{bc}}{(\bar{z}-\bar{\zeta})^2}+\frac{i\epsilon^{bcd}}{\bar{z}-\bar{\zeta}} J^d_R(\bar{\zeta})
\end{eqnarray}
Let $z=z_n$ and $\zeta=z_{n+1}$ such that $z_n-z_{n+1}=-ia$ and $\bar{z}_n-\bar{z}_{n+1}=ia$. Then the remaining $J_LJ_L$ and $J_RJ_R$ terms in the first line of Eq.~\eqref{3spin} simplify to
\begin{eqnarray}
\nonumber&& \epsilon^{abc}\left( J_L^b(z_{n}) J_L^c(z_{n+1}) +  J_R^b(\bar{z}_{n}) J_R^c(\bar{z}_{n+1}) \right) \\ \nonumber
&=& i\epsilon^{abc}\epsilon^{bcd}\left[\frac{1}{(-ia)}J^d_L(z_{n+1})+\frac{1}{ia}J^d_R(\bar{z}_{n+1}) \right] \\
\nonumber &=& \frac{2}{a}\left(J^b_R(\bar{z}_{n+1}) - J^b_L(z_{n+1})\right)
\end{eqnarray}

The $J_Lg$ and $J_Rg$ terms in the second and third lines of Eq.~\eqref{3spin} have OPE 
\begin{eqnarray}
\nonumber J^b_L(z) g(\zeta,\bar{\zeta}) &=& \frac{\sigma^b/2}{z-\zeta}g(\zeta,\bar{\zeta}) \\
\nonumber J^b_R(\bar{z}) g(\zeta,\bar{\zeta}) &=& -g(\zeta,\bar{\zeta}) \frac{\sigma^b/2}{\bar{z}-\bar{\zeta}}
\end{eqnarray}
for small $|z-\zeta|$ so in Eq.~\eqref{3spin} they reduce to
\begin{eqnarray}
\nonumber && \mbox{tr}(g(z_n,\bar{z}_n)\sigma^b)(J_L^c(z_{n+1})+J_R^c(\bar{z}_{n+1})) \\ \nonumber
&-&(J_L^b(z_n)+J_R^b(\bar{z}_n))\mbox{tr}(g(z_{n+1},\bar{z}_{n+1})\sigma^c) \\
\nonumber &=& \mbox{tr}\left(\left[\frac{\sigma^c}{2}\frac{1}{ia}g(z_n,\bar{z}_n)-g(z_n,\bar{z}_n)\frac{\sigma^c}{2}\frac{1}{(-ia)}\right]\sigma^b\right) \\
\nonumber &-& \mbox{tr}\left(\left[\frac{\sigma^b}{2}\frac{1}{(-ia)}g(z_{n+1},\bar{z}_{n+1})-g(z_{n+1},\bar{z}_{n+1})\frac{\sigma^b}{2}\frac{1}{ia}\right]\sigma^c\right) \\
\nonumber &=&\frac{1}{2ia}
\Bigl [
\mbox{tr}(g(z_n,\bar{z}_n)(\sigma^b\sigma^c+\sigma^c\sigma^b))  \\ \nonumber
 && \qquad  +  \mbox{tr}(g(z_{n+1},\bar{z}_{n+1})(\sigma^c\sigma^b+\sigma^b\sigma^c))
\Bigr ]
 \\
\nonumber &=&\frac{\delta^{bc}}{ia}\left[\mbox{tr}(g(z_n,\bar{z}_n))+\mbox{tr}(g(z_{n+1},\bar{z}_{n+1}))\right]
\end{eqnarray}
which is zero when contracted with the anti-symmetric Levi-Civita symbol. The $gg$ term in the last line of Eq.~\eqref{3spin} can be expanded in an OPE. To do so, we note that the $SU(2)$ matrix $g$ may be written in terms of a holomorphic and an anti-holomorphic function as $g^\alpha_{\mbox{ }\beta}(z,\bar{z})=g^\alpha(z)g_\beta(\bar{z})$.  The holomorphic and anti-holomorphic fields $g^\alpha(z)$ and $g_\beta(\bar{z})$ have OPE
\begin{eqnarray}
\nonumber g^{\alpha_1}(z)g^{\alpha_3}(\zeta) &=& \frac{\epsilon^{\alpha_1\alpha_3}}{\sqrt{z-\zeta}}+\sqrt{z-\zeta}\frac{(\sigma^c)^{\alpha_1\alpha_3}}{2}J^c_L(\zeta) \\
\nonumber g_{\alpha_2}(\bar{z})g_{\alpha_4}(\bar{\zeta}) &=& \frac{\epsilon_{\alpha_2\alpha_4}}{\sqrt{\bar{z}-\bar{\zeta}}}+\sqrt{\bar{z}-\bar{\zeta}}\frac{(\sigma^c)_{\alpha_2\alpha_4}}{2}J^c_R(\bar{\zeta})
\end{eqnarray}

Using this OPE, the last term in Eq.~\eqref{3spin} may be simplified to
\begin{eqnarray}
\nonumber && (\sigma^b)^{\alpha_2}_{\mbox{ }\alpha_1} (\sigma^c)^{\alpha_4}_{\mbox{ }\alpha_3}g^{\alpha_1}(z_n)g^{\alpha_3}(z_{n+1})g_{\alpha_2}(\bar{z}_n)g_{\alpha_4}(\bar{z}_{n+1}) \\
\nonumber &=& \frac{\epsilon^{\alpha_1\alpha_3}\epsilon_{\alpha_2\alpha_4}(\sigma^b)^{\alpha_2}_{\mbox{ }\alpha_1}(\sigma^c)^{\alpha_4}_{\mbox{ }\alpha_3}}{a} \\ \nonumber
&&+\frac{a}{4}(\sigma^b)^{\alpha_2}_{\mbox{ }\alpha_1} (\sigma^c)^{\alpha_4}_{\mbox{ }\alpha_3} (\sigma^d)^{\alpha_1\alpha_3}(\sigma^e)_{\alpha_2\alpha_4} J^d_L(z_{n+1}) J^e_R(\bar{z}_{n+1}) \\
\nonumber &+& \frac{1}{2}(-i)(\sigma^b)^{\alpha_2}_{\mbox{ }\alpha_1}(\sigma^c)^{\alpha_4}_{\mbox{ }\alpha_3}(\sigma^d)^{\alpha_1\alpha_3}\epsilon_{\alpha_2\alpha_4}J^d_L(z_{n+1}) \\ \nonumber
&&+  \frac{1}{2}(i)(\sigma^b)^{\alpha_2}_{\mbox{ }\alpha_1}(\sigma^c)^{\alpha_4}_{\mbox{ }\alpha_3}(\sigma^e)_{\alpha_2\alpha_4}\epsilon^{\alpha_1\alpha_3}J^e_R(\bar{z}_{n+1}) \\
\nonumber &=& \frac{2}{a}\delta^{bc}-\frac{a}{2}(\delta^{bd}\delta^{ec}+\delta^{be}\delta^{dc}-\delta^{bc}\delta^{de}) J^d_L(z_{n+1}) J^e_R(\bar{z}_{n+1}) \\ \nonumber
&&+ \epsilon^{bcd} J^d_L(z_{n+1}) -\epsilon^{bce} J^e_R(\bar{z}_{n+1})
\end{eqnarray}

The terms containing a $\delta^{bc}$ sum to zero when contracted with the Levi-Civita tensor. Therefore the last term in Eq.~\eqref{3spin} reduces to
\begin{eqnarray}
&& \epsilon^{abc}\mbox{tr}(g(z_{n},\bar{z}_{n})\sigma^b)\mbox{tr}(g(z_{n+1},\bar{z}_{n+1})\sigma^c) 
\\ \nonumber
&&= 2(J^a_L(z_{n+1})-J^a_R(\bar{z}_{n+1}))
\end{eqnarray}
Gathering all non-zero terms, the scalar spin chirality interaction becomes
\be
\epsilon^{abc} S^b_n S^c_{n+1} \rightarrow av(J^a_R(\bar{z})-J^a_L(z)) + O(a^3).
\ee
in the low energy continuum limit where $z=z_{n+1}$ and $v=2(1-a\times\mbox{const}^2)>0$ so we can set $av=K$.

%%%%%%%%%%%%%%%%%%%%%%%%%%%%%%%%%%%%%%%%%%%%%%%%%%%%%%%%%

%%%%%%%%%%%%%%%%%%%%%%%%%%%%%%%%%%%%%%%%%%%%%%%%%%%%%%%%%

\section{Majorana fermion model}
\label{app:MajoranaModel}

\begin{figure}
\centering
\includegraphics[width=\columnwidth]{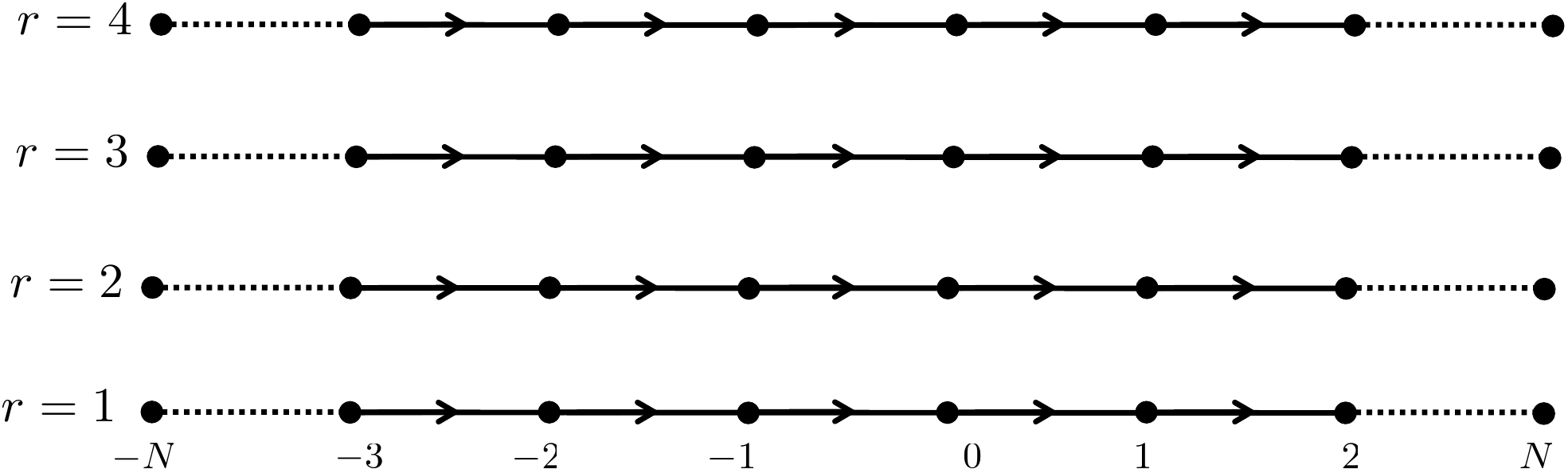}
\caption{A stack of $n=4$ free Majorana fermion chains. The arrows indicate the direction hopping where the hopping term is positive (i.e. $i\gamma_j\gamma_{j+1}$ as opposed to $-i\gamma_{j+1}\gamma_j$). All hoppings have positive  coupling strength $J>0$}
\label{MajoranaChainStack}
\end{figure}

We can carry out the weaving procedure
discussed in Sec.~\ref{sec:weaving}
 more explicitly in the quadratic  Majorana fermion model that we had already considered in section
\ref{SubSectionModelsAndLineOfArgument},
where we replaced the spin-$1/2$ operator at each lattice site by a Majorana fermion zero mode. We remind the reader (as in Sec.~\ref{SubSectionModelsAndLineOfArgument})
 that this
model is different from the spin model, 
but in this model we can infer the behavior  of the left- and right-movers more explicitly. Each spin chain in
Fig.~\ref{SpinChains1}
 is now replaced by a chain of Majorana fermion zero modes  (see Fig.~\ref{MajoranaChainStack}) with Hamiltonian
\begin{eqnarray}
\label{FermionChains} 
H = \sum_{r=1}^n\ \ H^r, \ \  \  
H^r \equiv i J \sum_{j=-N}^N ( \gamma^r_j \gamma^r_{j+1} + \eta^r_j \eta^r_{j+1}) \ \  \
\end{eqnarray}
where $J>0$,  $r$ labels the chains (vertical direction) and $j$ labels the sites along the chain (horizontal direction) with anti-periodic boundary conditions $\gamma_{N+1}^r=(-1)\gamma_{-N}^r$, and likewise for $\eta^r$. We found it convenient to double the Majorana fermion species to form ordinary complex fermions by introducing a replica $\eta^r_j$ of each Majorana fermion $\gamma^r_j$. The chemical potential of the resulting  model of complex fermions is zero corresponding to half-filling.

In each chain, this doubled Majorana
Hamiltonian is related through a gauge transformation to the more familiar one of a chain of complex fermions $c^r_j$ and ${c^r_j}^\dagger$ with nearest neighbor hopping,
i.e. $ H^r = J\sum_{j=-N}^{+N} ({c^r}^\dagger_j c^r_{j+1} + h.c.)$.
To see this, first consider forming complex fermions $a^r_j = (\gamma^r_j + i \eta^r_j)/\sqrt{2}$, $a^r_j = (\gamma^r_j - i \eta^r_j)/\sqrt{2}$. The conventional fermion operators
$c^r_j$ are then defined through the local gauge transformation $c^r_j=e^{i\frac{\pi}{2}j}a^r_j$. 
The annihilation and creation operators $c^r_j$ and ${c^r_j}^\dagger$ at site $j$ along chain $r$  may be written as the product of a rapidly oscillating phase $e^{ ix_j p_{L,R}}$ and a left- or right-moving fermion operator $\Psi^r_{L,R}(x_j)$ which varies only slowly  on the scale of the lattice spacing $a$, i.e. with $x_j=aj$,
\bea
\label{c1} c_j^r &=& e^{ix_jp_F}\Psi^r_R(x_j)+e^{-ix_jp_F}\Psi^r_L(x_j) \\
\label{c2} (c_j^r)^\dagger &=& e^{-ix_jp_F}(\Psi^r_R)^\dagger(x_j)+e^{ix_jp_F}(\Psi^r_L)^\dagger(x_j),
\eea
where half-filling fixes $p_F$ to $\pi/2a$.
We deduce from this the slowly varying Majorana degrees of freedom,
$\gamma_{L,R}(x_j)=(\Psi_{L,R}(x_j)+\Psi^\dagger_{L,R}(x_j))/\sqrt{2}$ and $\eta_{L,R}(x_j)=(\Psi_{L,R}(x_j)-\Psi^\dagger_{L,R}(x_j))/i\sqrt{2}$,
which are related to the lattice Majorana fermion operators via
\bea
\label{GammaContinuum} \gamma^r_j &=& (-1)^j \gamma^r_R(x_j) + \gamma^r_L(x_j) \\
\label{GammaContinuum2} \eta_j &=& (-1)^j \eta^r_R(x_j) + \eta^r_L(x_j).
\eea
As is well familiar, the Hamiltonian in Eq. (\ref{FermionChains}) reads when expressed in terms of the low-energy degrees of freedom
\begin{eqnarray}
\label{MajoranaHamiltonianLowEnergy} 
H^r  =v_F
\int {dx \over 2\pi}
[
\gamma^r_L(x) i {d\over dx} \gamma^r_L(x)
-
\gamma^r_R(x) i {d\over dx} \gamma^r_R(x)
]
 \ \  \
\end{eqnarray}
and changing the sign of $J$ exchanges the roles of  right- and left-movers.

\begin{figure}
\centering
\includegraphics[width=0.6\columnwidth]{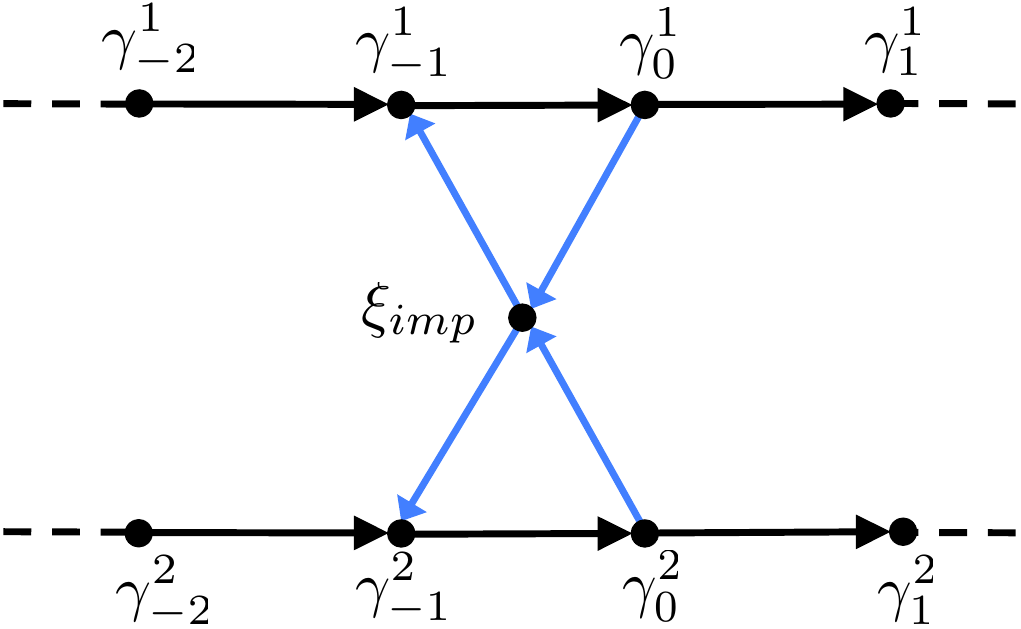}
\caption{The Majorana zero mode impurity $\xi_{imp}$ is coupled to two neighboring Majorana chains. The arrows indicate the chirality of the Majorana hoping operators. The black arrows have coupling strength $J$ and the blue arrows, corresponding to the interaction in Eq.~\eqref{FermionChiral}, have coupling strength $K$}
\label{MajoranaR}
\end{figure}

We now consider the case of 2 chains, say $r=1,2$, and add a single  interstitial
Majorana ``impurity zero mode'' $\xi_{\mathrm{imp}}$,
which we couple to the two closest Majorana fermion zero modes of each of the two adjacent
Majorana chains;
 see Fig.~\ref{MajoranaR}. This is in complete analogy to the (Kondo) spin case
[Fig.~\ref{SpinChains1}, and Eq. (\ref{SpinForm2channel})].
This coupling then reads
\bea
\label{FermionChiral}\delta H &=& iK \xi_{\mathrm{imp}}(\gamma^+_0-\gamma^+_{-1}) + (\gamma \leftrightarrow \eta)
\eea
where $\gamma^+_j$ corresponds to the symmetric combination $\gamma^+_j=\gamma^1_j+\gamma^2_j$ of the Majorana fermion operators on both chains;
see Fig.~\ref{MajoranaR} for the orientation of arrows. For later use we will also define an antisymmetric combination $\gamma_j^-=\gamma_j^1-\gamma_j^2$ which does not couple to $\xi_{\mathrm{imp}}$.
We perform the same steps for the $\eta$ degrees of freedom, including an independent ``impurity zero mode'' $\eta_{imp}$, but will for notational clarity not  write
these explicitily  in our discussion below.

Expressing the lattice fermions as in Eq.~\eqref{GammaContinuum} in terms of slowly varying fields $\gamma^\dagger_R$ and $\gamma^\dagger_L$, the interaction Eq.~\eqref{FermionChiral} with the impurity becomes
\bea
\label{Rcoupling}
 \delta H &=& 2K\xi_{\mathrm{imp}} \gamma^+_R(x_{-1}) + \mbox{RG irrel.}
\eea
where ``RG irrel.'' stands for terms that are irrelevant in the renormalization group (RG) sense. Because these operators are localized at one point in space, the RG irrelevant operators are those with a scaling dimension greater than $1$. Operators with 
derivatives are therefore always  RG irrelevant. Note that the only operator that appears in Eq.~\eqref{Rcoupling} that
is not RG irrelevant is the coupling
of the interstitial Majorana zero mode $\xi_{\mathrm{imp}}$ to the right-moving Majorana operator. This is a consequence of 
the chirality of the coupling in Eq.~\eqref{FermionChiral} [see Fig.~\ref{MajoranaR}], and of 
the minus sign in that same equation,
in combination with~\eqref{GammaContinuum}. This means that the impurity only couples to the right-moving Majorana degrees of freedom in the adjacent chains, and doesn't "see" the left movers (modulo irrelevant terms). The right-moving Majorana operators will interact with the impurity whereas the left-moving Majorana operators will remain decoupled.
We will now demonstrate that this gives
rise to the same configuration of left- and right-movers\footnote{To arrive at {\it precisely} the identical configuration, one should exchange
the role of left- and right-movers in the present Majorana fermion discussion, which is easily achieved by changing the sign of $J$  in Eq.
(\ref{FermionChains}) - see also Eq. (\ref{MajoranaHamiltonianLowEnergy}).}
as was seen previously for the more complicated 
spin model
 in Fig.~\ref{SpinChains1}.

\begin{figure}
\centering
\includegraphics[width=\columnwidth]{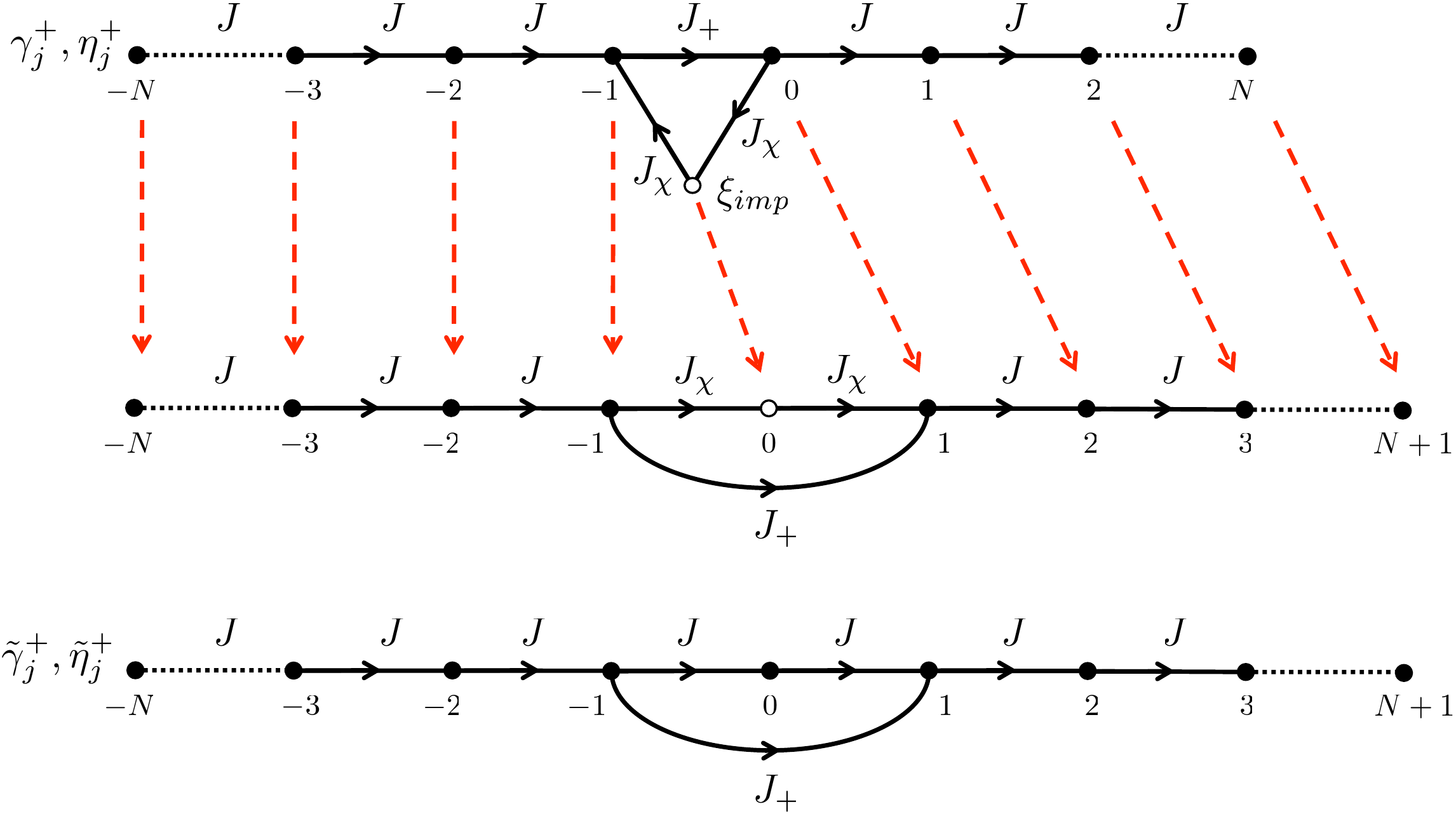}
\caption{By relabelling the sites of the $+$ Majorana fermion operators we see that the impurity $\xi_{imp}$ may be thought of as just another site along a chain of Majorana zero modes.}
\label{FermionRelabelling}
\end{figure}

\begin{figure}
\centering
\includegraphics[width=\columnwidth]{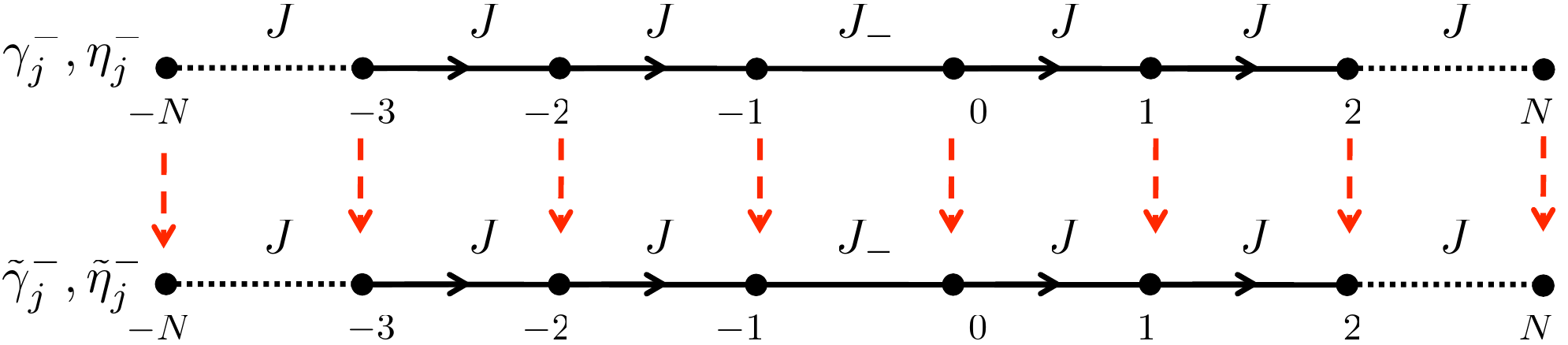}
\caption{The 
%$-$ 
chain of $-$ Majorana fermion operators doesn't couple to the impurity, and  therefore it remains unchanged under the transformation that adds the impurity as just another site along the 
%$-$ 
chain of $+$ Majorana fermion operators (see Fig.~\ref{FermionRelabelling})}
\label{MinusChain}
\end{figure}

The term written out in Eq.~\eqref{Rcoupling} is RG-relevant (with scaling dimension $=1/2$). In order to find the fate of this RG flow at large scales, we will now determine the infrared fixed point of this flow. As already mentioned, the left movers will be unaffected by the impurity whereas the right movers will interact with the impurity. This is analogous to the chiral Kondo effect discussed in Sec.~\ref{sec:technical}. We start by recalling that the impurity zero mode in Eq.~\eqref{FermionChiral} only interacts with the symmetric $+$ combination of the Majorana operators of both chains. We may therefore use the transformations $\gamma^{\pm}_j = \gamma^1_j\pm\gamma^2_j$ to rewrite our model, where the $+$ chain couples to the impurity and the $-$ chain doesn't couple. The $+$ chain and $-$ chain are shown in Fig.~\ref{FermionRelabelling} and Fig.~\ref{MinusChain}.

The interaction in (\ref{Rcoupling}) describes the well-known ``resonant level model''
 (which is exactly solvable by quadrature)~\cite{resonantlevelone,resonantleveltwo}.
Let us now consider the couplings $J_+$ and $J_{\chi}$ defined in Fig. \ref{FermionRelabelling}:
 When $J_+=0$, it is known that the 
coupling 
%$K$
$J_{\chi}$
 to the impurity ``heals'', and 
%$K$ 
$J_{\chi}$ thus flows to $J$ under renormalization group, where  
%$K=J$
$J_{\chi}=J$  denotes the stable infrared fixed point (the Majorana
analog of the 2-channel Kondo fixed point in the spin case).
By relabeling the sites in the $+$ chain as in the upper panel of Fig.~\ref{FermionRelabelling}, while still keeping $J_+ =0$, one
sees the emergence of the ``healed Majorana chain''.
So in the presence of the Majorana impurity zero mode,  the $+$ chain just absorbs the impurity as an extra ``interstitial'' site and the chain's length therefore increases by one site.

It turns out to be convenient to denote
the Majorana operators of the two  chains with one site inserted 
by $\tilde{\gamma}^{\pm}_j$. The transformations depicted  in Figs.~\ref{FermionRelabelling} and \ref{MinusChain} simply correspond to
\bea
\label{SiteTransformation1}\tilde{\gamma}^+_j &=& \gamma^+_j \mbox{ for } j<0 \\
\label{SiteTransformation2}
\tilde{\gamma}^+_0 &=& \xi_{\mathrm{imp}} \\
\label{SiteTransformation3} \tilde{\gamma}^+_{j+1} &=& \gamma^+_j \mbox{ for } j\geq 0 \\
\label{SiteTransformation4} \tilde{\gamma}^-_j &=& \gamma^-_j \mbox{ for all } j.
\eea
With the impurity now being part of the $+$ chain, the operators $\tilde{\gamma}^+_j$ and $\tilde{\gamma}^-_j$ of the $+$ and $-$ chains are just continuous across the site of the insertion. In particular, because the relabeled operators $\tilde{\gamma}^{\pm}_{L,R}(x_j)$ as  in Eq.~\eqref{GammaContinuum} are by definition slowly varying (continuous) on the scale of the lattice spacing, they satisfy the continuity conditions $\tilde{\gamma}^{\pm}_L(x_{-1})=\tilde{\gamma}^{\pm}_L(x_{+1})$ and $\tilde{\gamma}^{\pm}_R(x_{-1})=\tilde{\gamma}^{\pm}_R(x_{+1})$.

First, this is a good  
place to see that
 turning  on the coupling $J_+$ leads to an irrelevant operator, owing to reflection symmetry 
about the site of the chain which hosts the impurity: Considering
 the lower chain shown in the upper panel of Fig.~\ref{FermionRelabelling}, 
 the $J_+$-term
$J_+ i {\tilde \gamma}^+_{-1}{\tilde\gamma}^+_{+1}$  is immeditaly seen to only contain derivative operators
 when expressed in terms of the long wavelength degrees of freedom ${\tilde \gamma}^+_{L}$
and ${\tilde \gamma}^+_{R}$
using (\ref{GammaContinuum}).

Next, by inverting the  continuity conditions  Eqs.~\ref{SiteTransformation1}-\ref{SiteTransformation4}, we can transform them
back to the fermion degrees
of freedom of the original chains (those with Majorana operators $\gamma^{\pm}_j$).
This immediately yields the 
following continuity conditions of the left and right movers at the impurity site 
\bea
\label{BoundaryConditionsLeftMovers}
\gamma_L^+(x_{-1}) = \gamma_L^+(x_{0}), &\hspace{0.4cm}& \gamma_L^-(x_{-1}) = \gamma_L^-(x_0) \quad   \\
\label{BoundaryConditionsRightMovers}
\gamma_R^+(x_{-1}) = (-1)\gamma_R^+(x_0), &\hspace{0.4cm}& \gamma_R^-(x_{-1}) = \gamma_R^-(x_0) \quad.
\eea
The fields of the $-$ chain are 
continuous 
because they don't
couple to the impurity. The left-moving fermions of the $+$ chain
are also continous 
at the location of the impurity for the same reason. However, 
the right-moving fermion operators  of the $+$ chain acquire a minus sign as they go through the location of the impurity
Majorana zero mode. This sign change for the right movers of the $+$ chain
arises due to the combined effects of the site relabeling $j\rightarrow j+1$  (see Eq.~\eqref{SiteTransformation3}) after adding the impurity to the $+$ chain and because of the dependence of the sign of the coefficient of the right movers (see Eq.~\eqref{GammaContinuum}) on the lattice site $j$.

The $+$ and $-$ chains are just a convenient way for
obtaining these continuity conditions and we would like to rewrite them back in terms of the physical degrees of freedom of Fig.~\ref{MajoranaR}. To do this we invert the transformations $\gamma_j^{\pm}=\gamma_j^1\pm\gamma_j^2$ used previously. In terms of the Majorana operators on the upper $r=1$ and lower $r=2$ chain the continuity conditions then read 
\begin{align}
\gamma^1_R(x_{-1}) &= (-1)\gamma^2_R(x_0),   \label{KondoBC1}
&\gamma^2_R(x_{-1}) &= (-1)\gamma^1_R(x_0) \\
\gamma^1_L(x_{-1}) &= \gamma^1_L(x_0),
& \label{KondoBC4} \gamma^2_L(x_{-1}) &= \gamma^2_L(x_0).
\end{align}
For the left movers these are just free fermion continuity conditions. The left movers are therefore unaffected by the impurity as expected
(in the same way as depicted for right-movers in
 Fig.~\ref{SpinChains1}).
However the boundary conditions in Eq. (\ref{KondoBC1})
 for the right movers reveal that 
the right moving Majorana degrees of freedom get exchanged (up to a minus sign) upon crossing the location of the impurity.
For example the right mover on site $-1$ on the upper chain "jumps" to site $0$ on the lower chain across the impurity. Likewise the right mover on site $-1$ of the lower chain "jumps" to site $0$ on the upper chain as it crosses the impurity. These continuity conditions are therefore analogous to those discussed for the spin model in the context of
Fig.~\ref{SpinChains1}, and lead to the weaving construction that was discussed  in Sec. \ref{sec:weaving}.
We thus see that our Majorana model
realizes a Majorana (resonant level) version of the chiral two-channel Kondo effect discussed in Sec.~\ref{sec:weaving}.
This is the basis of the argument in Sec.~\ref{SubSectionValidationForTheMajoranaModel}.

We conclude this appendix by discussing the effect of possible exactly  marginal (boundary) operators located at the impurity site (there are no 
relevant  nor marginally relevant boundary operators, except for the interaction in (\ref{Rcoupling}) which we have already discussed).

\begin{figure}
\centering
\includegraphics[width=0.35\columnwidth]{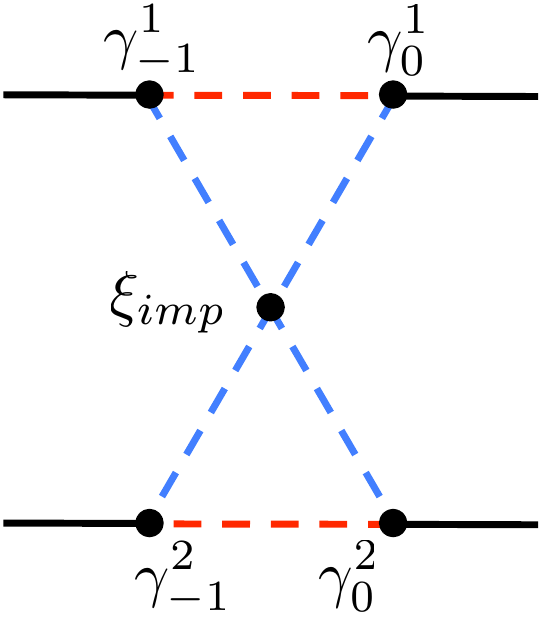}
\caption{At weak coupling $J_\chi$ there are two exactly marginal (boundary) operators located at the impurity site: the link coupling
$\delta H_{=}$ shown in red, and the cross-chain couping $\delta H_{X}$.}
\label{MajoranaMarginalImpurityInteractions}
\end{figure}

 In order for the impurity model to sit at the infrared
 fixed point described above, these marginal operators must be absent. Referring to 
Fig.~\ref{MajoranaMarginalImpurityInteractions},
one of these marginal operators $\delta H_{=}$ couples the two Majorana zero modes at site $-1$ and at site $0$ on the upper chain plus the two Majorana zero modes at site $-1$ and at site $0$ on the lower chain (red dashed line in 
Fig.~\ref{MajoranaMarginalImpurityInteractions}). The other marginal operator $\delta H_{\times}$ couples the Majorana zero mode at site $-1$ on the upper chain to the Majorana zero mode at site $0$ on the lower chain plus the Majorana zero mode at site $-1$ on the lower chain to the Majorana zero mode at site $0$ on the upper chain (blue dashed line in 
Fig.~\ref{MajoranaMarginalImpurityInteractions}).

By using the identity Eq.~\eqref{GammaContinuum} and the transformation to the $+$ and $-$ chain $\gamma^{\pm}_j=(\gamma^1_j\pm\gamma^2_j)$,
 we may reduce these 
operators to an interaction between left and right moving Majorana fermions
\bea
\label{MarginalLink} &&\delta H_{=} = J_{=} i (\gamma^1_j\gamma^1_{j+1} + \gamma^2_j \gamma^2_{j+1}) \\
\label{LinkLR} &&= J_{=} i(-1)^j2\left[\gamma^+_R(x_j)\gamma^+_L(x_j)+\gamma^-_R(x_j)\gamma^-_L(x_j)\right] \qquad
\\ &&+ \mbox{ derivatives},  \qquad \\
\label{MarginalCross} &&\delta H_{\times} = J_{\times} i (\gamma^1_j\gamma^2_{j+1} + \gamma^2_j \gamma^1_{j+1}) \\
\label{CrossLR} &&= J_{\times} i(-1)^j2\left[\gamma^+_R(x_j)\gamma^+_L(x_j)-\gamma^-_R(x_j)\gamma^-_L(x_j)\right] \qquad
\\ && + \mbox{ derivatives}, \qquad 
\eea
where use was made of Eq.~\eqref{GammaContinuum} and Eq.~\eqref{GammaContinuum2}. Therefore, in the low energy theory,
there are two exactly marginal operators,
one 
on the $+$ chain,
 $\delta H_{+}=J_{+}i(-1)^j \gamma^+_R(x_j)\gamma^+_L(x_j)$,
and other one
on the $-$ chain, 
$\delta H_{-}=2J_{-}i(-1)^j \gamma^-_R(x_j)\gamma^-_L(x_j)$,
where
$
J_{\pm}
= (J_{=}\pm J_{\times})$.
In the Hamiltonian of our impurity model these marginal operators disappear as follows. Referring to Fig.~\ref{MajoranaR}, the marginal operator of the $+$ chain ceases to exist (has become irrrelevant) at the strong coupling fixed point arising after the impurity Majorana fermion zero mode $\xi_{\mathrm{imp}}$ has become embedded into the $+$ chain. (That marginal operator is, in the new chain, odd under site parity about the location
of  $\xi_{\mathrm{imp}}$). On the other hand, the other marginal operator in the $-$ chain continues to be present. However, we are interested in the model where we have in the original formulation a coupling $J$ in the chains and a $3-$spin interaction $K$ around each triangle before going to the $+$ and $-$ fermion variables. This means that
$J_{=} = J+K$ and $J_{\times}=K$, implying that 
 in the $+$ and $-$ variables, $J_+= J+2K$, $J_-=J$. As we have seen, we can ignore $J_+$, but not $J_-$. However, as we go to the fully packed limit as in 
Fig.~\ref{SpinChains3},
we want to let $J\rightarrow 0$ and have the coupling $K$ take over the role of the original $J$. In this limit, clearly, the effective coupling  $J_-=J$
$\to 0$  disappears. In conclusion, we found that in the limit $J\to 0$ and $K=$ finite, studied in Sec.~\ref{SubSectionKagomeModel}, in 
the thermodynamic limit in the bulk, there are no marginal impurity operators appearing in the weaving construction. Such operators would appear
to be relevant in the limit of a regular array of interstitial Majorana impurity zero modes as in Fig.~\ref{SpinChains3}.

%%%%%%%%%%%%%%%%%%%%%%%%%%%%%%%%%%%%%%%%%%%%%%%%%%%%%%%%%

\section{Details of the 3D lattices}
\label{app:lattices}

\begin{figure}[h]
  \includegraphics[width=\columnwidth]{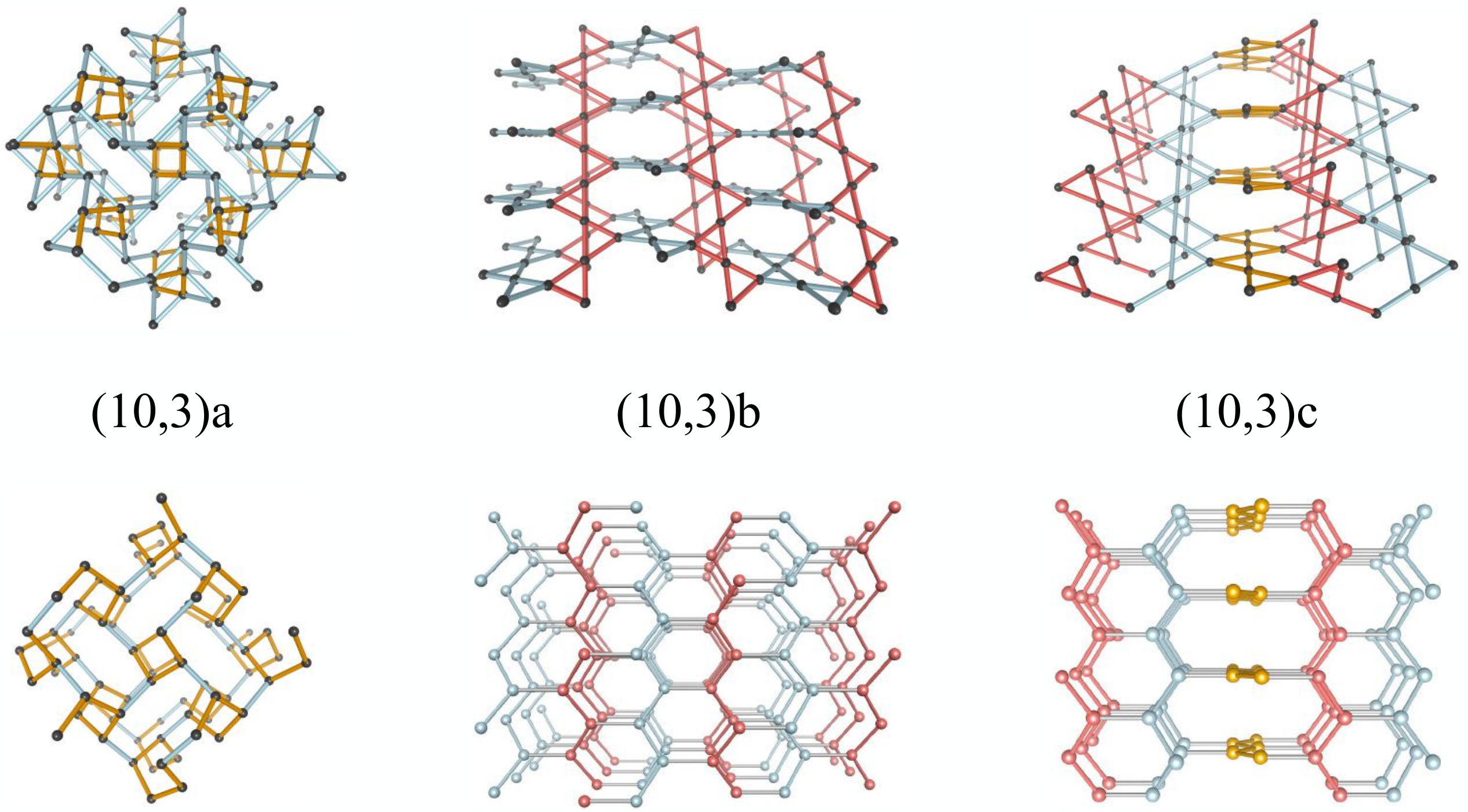}
  \caption{{\bf Overview of 3D lattices.}  
  		The top row shows three distinct three-dimensional lattices of corner-sharing triangles, along with their
		medial tricoordinated lattices in the bottom row. According to their classification in A.~F.~Wells' book 	
		\cite{Wells1977} these lattices are denoted by the Schl\"afli symbols as (10,3)a, (10,3)b, 
		and (10,3)c, respectively.
		The color-coding highlights the essential features of these lattices:
		Lattice (10,3)a consists of two counterrotating helices of length 4 (brown bonds) and length 8 (blue bonds).
		Lattices (10,3)b and (10,3)c consist of two/three zig-zag chains indicated in blue/red/yellow, with
		relative angles of 90/60 degrees, respectively.
  \label{fig:3D_Lattices} }
\end{figure}

In this appendix we provide detailed lattice information for the three-dimensional lattices of corner-sharing triangles 
that underly the 3D generalizations of Sec.~\ref{sec:3d}. 
For each lattice, we first provide information for the parent {\em tricoordinated} lattice, which we denote following
the notation of A.~F.~Wells' book \cite{Wells1977} by Schl\"afli symbols as (10,3)a, (10,3)b, and (10,3)c, respectively. 
For each tricoordinated lattice we subsequently provide information for its premedial lattice of corner-sharing triangles
relevant to the study at hand. A visual overview of these lattices is given in Fig.~\ref{fig:3D_Lattices}.

\subsection{Premedial lattice of (10,3)a}

The (10,3)a lattice is a body-centered-cubic lattice with four sites per unit cell at positions
\begin{align}
  \br_1&=\left(\frac 1 8, \frac 1 8 , \frac 1 8 \right) \,, &\br_2&=\left(\frac 5 8, \frac 3 8 , -\frac 1 8 \right) \,,\nonumber\\
  \br_3&=\left(\frac 3 8, \frac 1 8 , -\frac 1 8 \right) \,, &\br_4&=\left(\frac 7 8, \frac 3 8 , \frac 1 8 \right) \,.
\end{align}
The lattice vectors are given by 
\begin{align}
  \ba_1&=\left(1,0,0\right),&
  \ba_2&=\left(\frac{1}{2},\frac{1}{2},-\frac{1}{2}\right),&
  \ba_3&=\left(\frac{1}{2},\frac{1}{2},\frac{1}{2}\right),
\end{align}
and their corresponding reciprocal lattice vectors are
\begin{align}
  \bq_1 &= \left( 2\pi, -2\pi, 0 \right),&
  \bq_2 &= \left( 0, 2\pi, -2\pi \right),&\nonumber \\
  \bq_3 &= \left( 0, 2\pi, 2\pi \right).
  \label{eq:recip_vec_10_3a}
\end{align}
Its premedial lattice of corner-sharing triangles has six sites per unit cell at the positions 
\begin{align}
    \br_1 &=     \left(\frac{1}{8}, 0,  \frac{1}{4}\right) \,,
    & \br_2 &=    \left(\frac{3}{8}, 0,  -\frac{1}{4}\right) \,, \nonumber\\ 
    \br_3 &=     \left(\frac{1}{4},  \frac{1}{8}, 0 \right) \,,
    & \br_4 &=     \left(\frac{3}{4},  \frac{3}{8}, 0 \right) \,, \nonumber\\
    \br_5 &=    \left(\frac{1}{2},  \frac{1}{4}, -\frac{1}{8} \right) \,,
    & \br_6 &=    \left(0,  \frac{1}{4}, \frac{1}{8} \right) \,. 
\end{align}

\subsection{Premedial lattice of (10,3)b}

The tricoordinated (10,3)b lattice is a tetragonal lattice with four sites per unit cell at positions
\begin{align}
  \br_1& = \left( \frac{1}{2}, \frac{\sqrt{3}}{10} , 0 \right)\,, &\br_2& = \left( \frac{3}{5}, \frac{\sqrt{3}}{5} , \frac{2\sqrt{2}}{5} \right) \,,\nonumber\\
  \br_3 &= \left( \frac{2}{5}, \frac{\sqrt{3}}{5} , \frac{\sqrt{2}}{5} \right)\,, &\br_4 &= \left( 0, \frac{2\sqrt{3}}{5} , 0 \right)\,.
\end{align}
The lattice vectors are given by 
\begin{align}
  \ba_1 &= \left( -1, 1, -2 \right),&
  \ba_2 &= \left( -1, 1, 2 \right),&
  \ba_3 &= \left( 2, 4, 0 \right),
\end{align}
and their corresponding reciprocal lattice vectors are
\begin{align}
  \bq_1 &= \left( -\frac{2\pi}{3}, \frac{\pi}{3}, -\frac{\pi}{2}  \right),&
  \bq_2 &= \left( -\frac{2\pi}{3}, \frac{\pi}{3}, \frac{\pi}{2}  \right),&\nonumber \\
  \bq_3 &= \left( \frac{\pi}{3}, \frac{\pi}{3}, 0 \right).
  \label{eq:recip_vec_10_3b}
\end{align}
Its premedial lattice of corner-sharing triangles has six sites per unit cell at the positions 
\begin{align}
    \br_1 &=     \left(1, \frac{3}{2}, \frac{1}{2} \right) \,,
    & \br_2 &=    \left(2, \frac{7}{2}, \frac{1}{2} \right) \,, \nonumber\\ 
    \br_3 &=     \left(\frac{5}{2}, 3, \frac{3}{2} \right) \,,
    & \br_4 &=     \left(\frac{1}{2}, 2, \frac{3}{2} \right) \,, \nonumber\\
    \br_5 &=    \left(\frac{1}{2},  \frac{1}{2}, 0 \right) \,,
    & \br_6 &=    \left(\frac{3}{2}, \frac{5}{2}, 1 \right) \,. 
\end{align}

\subsection{Premedial lattice of (10,3)c}

The tricoordinated lattice (10,3)c is a trigonal lattice with six sites per unit cell at positions
\begin{align}
  \br_1 &= \left( \frac{1}{4}, \frac{1}{4\sqrt{3}}, \frac{1}{2\sqrt{3}} \right)\,, &\br_2 &= \left( \frac{3}{4}, \frac{1}{4\sqrt{3}}, \frac{2}{\sqrt{3}} \right)\,, \nonumber\\ 
  \br_3 &= \left( \frac{1}{2}, \frac{1}{\sqrt{3}}, \frac{7}{2\sqrt{3}} \right)\,, &\br_4 &= \left( \frac{3}{4}, \frac{1}{4\sqrt{3}}, \frac{1}{\sqrt{3}} \right)\,, \nonumber\\ 
  \br_5 &= \left( \frac{1}{2}, \frac{1}{\sqrt{3}}, \frac{5}{2\sqrt{3}} \right)\,, &\br_6 &= \left( \frac{1}{4}, \frac{1}{4\sqrt{3}}, \frac{4}{\sqrt{3}} \right)\,. 
\end{align}
The lattice vectors are given by 
\begin{align}
  \ba_1 &= \left( 1, 0, 0 \right),&
  \ba_2 &= \left( -\frac{1}{2}, \frac{\sqrt{3}}{2}, 0 \right),&\nonumber \\
  \ba_3 &= \left( 0, 0, \frac{3\sqrt{3}}{2} \right),
\end{align}
and their corresponding reciprocal lattice vectors are
\begin{align}
  \bq_1 &= \left( 2\pi, \frac{2\pi}{\sqrt{3}}, 0 \right),&
  \bq_2 &= \left( 0, \frac{4\pi}{\sqrt{3}}, 0 \right),&\nonumber \\
  \bq_3 &= \left( 0, 0, \frac{4\pi}{3\sqrt{3}} \right).
  \label{eq:recip_vec_10_3c}
\end{align}

Its premedial lattice of corner-sharing triangles has nine sites per unit cell at the positions 
\begin{align}
    \br_1 &=     \left(0, \frac{1}{4\sqrt{3}}, \frac{\sqrt{3}}{4}\right)\,,
    & \br_2 &=     \left(\frac{1}{2}, \frac{1}{4\sqrt{3}}, \frac{\sqrt{3}}{4}\right) \,, \nonumber\\ 
    \br_3 &=     \left(\frac{3}{4}, \frac{1}{4\sqrt{3}}, \frac{\sqrt{3}}{2}\right) \,,
    & \br_4 &=     \left(\frac{5}{8}, \frac{5}{8\sqrt{3}}, \frac{3\sqrt{3}}{4}\right) \,, \nonumber\\
    \br_5 &=     \left(\frac{3}{8}, \frac{11}{8\sqrt{3}}, \frac{3\sqrt{3}}{4}\right) \,,
    & \br_6 &=     \left(\frac{1}{2}, \frac{1}{\sqrt{3}}, \sqrt{3}\right) \,, \nonumber\\
    \br_7 &=     \left(\frac{5}{8}, \frac{11}{8\sqrt{3}}, \frac{5\sqrt{3}}{4}\right) \,,
    & \br_8 &=     \left(\frac{3}{8}, \frac{5}{8\sqrt{3}}, \frac{5\sqrt{3}}{4}\right) \,, \nonumber\\
    \br_9 &=     \left(\frac{1}{4}, \frac{1}{4\sqrt{3}}, \frac{3\sqrt{3}}{2}\right) \,.
\end{align}
%

%%%%%%%%%%%%%%%%%%%%%%%%%%%%%%%%%%%%%%%%%%%%%%%%%%%%%%%%%
\begin{widetext}
\section{Hyperkagome model}
\label{app:hyperkagome}

In this Appendix, we first provide additional details for the argument discussed in Sec.~\ref{sec:symm_even}, and then proceed to apply this line of argument
to the three-dimensional hyperkagome lattice, where we show that a Hamiltonian of the form~\eqnref{eqn:MajH} anticommutes with lattice symmetries which
guarantee gapless lines.

\subsection{Supporting calculations}

We first review the derivation of Eqn.~\eqnref{eqn:gammasymmaction}. We start from real-space Majorana operators $\gamma_{\alpha}(\vec{R})$, where
$\vec{R}$ is the origin of the unit cell and the sublattice site is offset by $\vec{s}_\alpha$ from the origin. Under the lattice symmetry $\Ss$, these transform as
\begin{equation}
\Ss \gamma_\alpha (\vec{R}) \Ss = \gamma_{\check{\alpha}}(\check{\vec{R}}+\vec{o}_\alpha).
\end{equation}
We again use the notation that $\Ss(\vec{x}) = \check{\vec{x}}+\vec{v}$ and denote as $\check{\alpha}$ the site in the unit cell into which the $\alpha$'th site
is mapped under the symmetry, and we will use that $\check{\vec{v}} \cdot \vec{w} = \vec{v} \cdot \check{\vec{w}}$.
For the momentum-space operators, we make the ansatz
\begin{equation} \label{eqn:app-kansatz}
\gamma_\alpha (\vec{k}) = \sum_{\vec{R}} e^{i \vec{k} \cdot (\vec{R}+\vec{s}_\alpha)} \gamma_\alpha(\vec{R}).
\end{equation}
We find that the action of $\Ss$ on these is given by (cf Eqn.~\eqnref{eqn:gammasymmaction}):
\begin{align}
\Ss \gamma_\alpha(\vec{k}) \Ss &= \sum_{\vec{R}} e^{i \vec{k} \cdot (\vec{R}+\vec{s}_\alpha)} \Ss \gamma_\alpha(\vec{R}) \Ss \nonumber \\
&= \sum e^{i \vec{k} \cdot (\vec{R}+\vec{s}_\alpha)} \gamma_{\check{\alpha}}(\check{\vec{R}}+\vec{o}_\alpha) \nonumber \\
&= \sum e^{i \vec{k} \cdot (\check{\vec{R}}+\vec{s}_\alpha-\check{\vec{o}}_\alpha)} \gamma_{\check{\alpha}}(\vec{R}) \nonumber \\
&= \sum e^{i \check{\vec{k}} \cdot (\vec{R}+\check{\vec{s}}_\alpha-\vec{o}_\alpha)} \gamma_{\check{\alpha}}(\vec{R}) \nonumber\\
&= e^{i \check{\vec{k}} \cdot (\check{\vec{s}}_\alpha-\vec{o}_\alpha)} e^{-i \check{\vec{k}} \cdot \vec{s}_{\check{\alpha}}} \sum e^{i \check{\vec{k}} \cdot (\vec{R}+\vec{s}_{\check{\alpha}})} \gamma_{\check{\alpha}}(\vec{R}) \nonumber \\
&= e^{i \check{\vec{k}} \cdot (\check{\vec{s}}_\alpha-\vec{o}_\alpha-\vec{s}_{\check{\alpha}})} \gamma_{\check{\alpha}}(\check{\vec{k}}) \,.
\end{align}

\subsection{A simple warmup case}

Let's reconsider the example of Sec.~\ref{sec:ff1d} of a one-dimensional lattice with two sites in the unit cell and an inversion symmetry about one of the sites.
For consistency across this section, we first repeat the example of Sec.~\ref{sec:ff1d} in slightly different notation, and then present a different example where
instead of a site inversion symmetry, we consider a bond inversion symmetry, which we find to not lead to any constraints.

Consider the operators $\gamma_0$ at $s_0=0$, and $\gamma_1$ and $s_1 = 1/2$, and the symmetry operation of a reflection about the $\gamma_1$ operators
in the unit cell at $R=0$, $\Ss(x) = -x+1$ (and thus $\check{x} = -x$). Then we have that
\begin{align}
\Ss \gamma_0(R) \Ss &= \gamma_0(-R+1) & \Ss \gamma_1(R) \Ss &= \gamma_1(-R)
\end{align}
To see the second case, consider that $\gamma_1(R)$ is at position $R+s_1 = R+1/2$, which under $\Ss$ goes to $-R+1/2$. The offset vectors are then
$o_0 = 1$ and $o_1 = 0$. For $\gamma_0$, using $\check{s}_0=0$, $s_{\check{0}} = 0$, we find the momentum-space operators to transform as 
\begin{equation}
\Ss \gamma_0(k) \Ss = e^{i (-k) (0-1-0)} \gamma_0(-k) = e^{ik} \gamma_0(-k).
\end{equation}
Similarly, using $\check{s}_1 = -1/2$, $s_{\check{1}} = 1/2$, and $o_1 = 0$, we have
\begin{equation}
\Ss \gamma_1(k) \Ss = e^{i (-k) (-1/2 - 0 -1/2)} \gamma_1(-k) = e^{ik} \gamma_1(-k).
\end{equation}
If we now consider a momentum that is invariant under the symmetry, $k=-k=0$, we find that $U = \id$ and thus has signature $+2$, implying that the Hamiltonian
vanishes at $k=0$.

An interesting alternative is to consider the reflection about the center of the bond between the $\gamma_0$ and $\gamma_1$ operators, $\Ss'(x) = -x + 1/2$. Then,
\begin{align}
\Ss' \gamma_0(R) \Ss' &= \gamma_1(-R) & \Ss' \gamma_1(R) \Ss' &= \gamma_0(R)
\end{align}
and therefore $o_0=o_1=0$. Using $\check{s}_0=0$, $s_{\check{0}} = 1/2$, $\check{s}_1 = -1/2$ and $s_{\check{1}} = 0$, we find for the momentum-space operators
\begin{align}
\Ss' \gamma_0(k) \Ss' &= e^{i k/2} \gamma_1(-k) & \Ss' \gamma_1(k) \Ss' = e^{i k/2} \gamma_0(-k)
\end{align}
and thus $U(k) = e^{i k/2} \sigma_x$, where $\sigma_x$ is the usual Pauli matrix. At $k=0$, it follows that $U=\sigma_x$, which has signature zero
and therefore does not enforce any gapless modes at this momentum. Indeed, under these symmetry constraints it is easy to write down a fully gapped Hamiltonian:
$H = \sum_R i \gamma_0(R) \gamma_1(R)$.

\subsection{Hyperkagome case}

We now discuss in detail the case of the hyperkagome lattice, where the results of Sec.~\ref{sec:symm_even} will be used to show that
gapless lines in momentum space are protected by the lattice symmetries.
The hyperkagome lattice~\cite{Okamoto2007} is a three-dimensional network of cornersharing triangles forming a chiral cubic structure.
With the cubic lattice spanned by the basis vectors
\begin{align}
a_1 = \left( \begin{array}{c} 1\\0\\0 \end{array} \right) \ \ 
a_2 = \left( \begin{array}{c} 0\\1\\0 \end{array} \right) \ \ 
a_3 = \left( \begin{array}{c} 0\\0\\1 \end{array} \right) \,,
\end{align}
we choose a unit cell consisting of 12 sites, which are
\begin{subequations}
\begin{align}s_{0} &= \left(1/8, 1/8, -1/8 \right)^T&s_{1} &= \left(3/8, -1/8, 3/8 \right)^T&s_{2} &= \left(-1/8, 5/8, 5/8 \right)^T\\ s_{3} &= \left(5/8, 3/8, 1/8 \right)^T&s_{4} &= \left(-1/8, 1/8, 1/8 \right)^T&s_{5} &= \left(3/8, 3/8, -1/8 \right)^T\\ s_{6} &= \left(5/8, -1/8, 5/8 \right)^T&s_{7} &= \left(1/8, 5/8, 3/8 \right)^T&s_{8} &= \left(1/8, -1/8, 1/8 \right)^T\\ s_{9} &= \left(-1/8, 3/8, 3/8 \right)^T&s_{10} &= \left(5/8, 5/8, -1/8 \right)^T&s_{11} &= \left(3/8, 1/8, 5/8 \right)^T\end{align}
\end{subequations}

To define our model, we consider triangles with vertices $(v_1,v_2,v_3)$ formed by the lattice sites
\begin{equation}
\begin{array}{|l|l|l|} \hline
v_1 &v_2 &v_3 \\ \hline \hline s_{0}& s_{8}& s_{4}\\ \hline s_{0}& s_{11} + \left(0,0,-1 \right)^T& s_{5}\\ \hline s_{1}& s_{11}& s_{6}\\ \hline s_{1}& s_{8}& s_{7} + \left(0,-1,0 \right)^T\\ \hline s_{2}& s_{9}& s_{7}\\ \hline s_{2}& s_{10} + \left(-1,0,1 \right)^T& s_{6} + \left(-1,1,0 \right)^T\\ \hline s_{3}& s_{10}& s_{5}\\ \hline s_{3}& s_{9} + \left(1,0,0 \right)^T& s_{4} + \left(1,0,0 \right)^T\\ \hline \end{array}
\end{equation}
Note that the ordering of vertices on each triangles matters for the chiral Hamiltonian. In principle, the sign of the coupling for each triangle can be chosen independently, leading to
$2^{8-1}$ possible Hamiltonians (since the overall sign does not matter). The particular choice we make is encoded in the ordering of vertices above, and is chosen in such a way that
wherever two triangles meet and two of their edges are parallel, the chirality along these edges coincides.
The spin and Majorana
Hamiltonians are then given by, respectively,
\begin{eqnarray}
H &=& \sum_{v_1, v_2, v_3 \in \bigtriangleup} S_{v_1} \cdot \left( S_{v_2} \times S_{v3} \right) \\
\tilde{H} &=& i \sum_{v_1, v_2, v_3 \in \bigtriangleup} \left( \gamma_{v_1} \gamma_{v_2} + \gamma_{v_2} \gamma_{v_3} + \gamma_{v_3} \gamma_{v_1} \right).
\end{eqnarray}

We now focus on the case of free Majorana fermions, and use the ansatz of Eqn.~\eqnref{eqn:kansatz} to bring the Hamiltonian into the form of Eqn.~\eqnref{eqn:hk}
with $h(k) = b(k) + b^\dagger(k)$, and
\begin{equation}
b(k) = i\left(
\begin{array}{cccccccccccc}0& 0& 0& 0& 0& 0& 0& 0& e^{ i k \cdot v_{0}} & 0& 0& e^{ i k \cdot v_{2}}  \\ 0& 0& 0& 0& 0& 0& 0& 0& e^{ i k \cdot v_{5}} & 0& 0& e^{ i k \cdot v_{3}}  \\ 0& 0& 0& 0& 0& 0& 0& 0& 0& e^{- i k \cdot v_{3}} & e^{- i k \cdot v_{2}} & 0 \\ 0& 0& 0& 0& 0& 0& 0& 0& 0& e^{- i k \cdot v_{5}} & e^{- i k \cdot v_{0}} & 0 \\ e^{ i k \cdot v_{2}} & 0& 0& e^{ i k \cdot v_{1}} & 0& 0& 0& 0& 0& 0& 0& 0 \\ e^{ i k \cdot v_{4}} & 0& 0& e^{- i k \cdot v_{5}} & 0& 0& 0& 0& 0& 0& 0& 0 \\ 0& e^{ i k \cdot v_{5}} & e^{- i k \cdot v_{1}} & 0& 0& 0& 0& 0& 0& 0& 0& 0 \\ 0& e^{- i k \cdot v_{4}} & e^{- i k \cdot v_{2}} & 0& 0& 0& 0& 0& 0& 0& 0& 0 \\ 0& 0& 0& 0& e^{ i k \cdot v_{1}} & 0& 0& e^{ i k \cdot v_{0}} & 0& 0& 0& 0 \\ 0& 0& 0& 0& e^{- i k \cdot v_{3}} & 0& 0& e^{- i k \cdot v_{4}} & 0& 0& 0& 0 \\ 0& 0& 0& 0& 0& e^{ i k \cdot v_{4}} & e^{- i k \cdot v_{0}} & 0& 0& 0& 0& 0 \\ 0& 0& 0& 0& 0& e^{ i k \cdot v_{3}} & e^{- i k \cdot v_{1}} & 0& 0& 0& 0& 0 \\ \end{array}
\right)
\end{equation}
where
\begin{subequations}
\begin{align}v_{0} &= (0,-1,1)^T/4&v_{1} &= (-1,1,0)^T/4&v_{2} &= (1,0,-1)^T/4\\ v_{3} &= (0,1,1)^T/4&v_{4} &= (-1,-1,0)^T/4&v_{5} &= (-1,0,-1)^T/4.\end{align}
\end{subequations}

We consider $\pi$ rotations around axis that lie in the planes formed by pairs of triangles. These planes are perpendicular to the four diagonals of the unit cell:
\begin{align}
v_1 &= (1,1,1) &
v_2 &= (-1,-1,1) \nonumber \\
v_3 &= (-1,1,1) &
v_4 &= (1,-1,1) \,.
\end{align}
For each diagonal, the three vectors orthogonal to it that correctly lie in the planes formed by the triangles are:
\begin{align}
v_1 &: (0,-1,1), (-1,0,1), (1,-1,0) \nonumber \\
v_2 &: (1,-1,0), (0,1,1), (1,0,1) \nonumber \\
v_3 &: (1,1,0), (1,0,1), (0,-1,1) \nonumber \\
v_4 &: (1,1,0), (-1,0,1), (0,1,1) \,.
\end{align}
Note that, as expected, the angles between any pair of vectors for a given $v_i$ are $2\pi/3$. Note that many of these coincide; the unique vectors are given
by
\begin{align}
w_1&=(0,-1,1) &w_2&=(-1,0,1) &w_3&=(1,-1,0) \nonumber \\
w_4&=(0,1,1) &w_5&=(1,0,1) &w_6&=(1,1,0) \,.
\end{align}
We now consider the $\pi$ rotations around these, and for each find an appropriate shift vector such that the combined action of the rotation and a
translation form a symmetry of the lattice. They are given by:
\begin{align} 
R_{w_1}(v) &=  \left( \begin{array}{ccc}  -1  &  0  &  0  \\  0  &  0  &  -1  \\  0  &  -1  &  0  \\  \end{array} \right) v  + \left( \begin{array}{c} 3/4 \\ 3/4 \\ -1/4 \end{array} \right)  \ \ \ \
& R_{w_2}(v) &=  \left( \begin{array}{ccc}  0  &  0  &  -1  \\  0  &  -1  &  0  \\  -1  &  0  &  0  \\  \end{array} \right) v  + \left( \begin{array}{c} 3/4 \\ 3/4 \\ -1/4 \end{array} \right)  \nonumber \\ 
R_{w_3}(v) &=  \left( \begin{array}{ccc}  0  &  -1  &  0  \\  -1  &  0  &  0  \\  0  &  0  &  -1  \\  \end{array} \right) v  + \left( \begin{array}{c} 3/4 \\ 3/4 \\ -1/4 \end{array} \right)  \ \ \ \
& R_{w_4}(v) &=  \left( \begin{array}{ccc}  -1  &  0  &  0  \\  0  &  0  &  1  \\  0  &  1  &  0  \\  \end{array} \right) v  + \left( \begin{array}{c} 1/4 \\ -1/4 \\ -3/4 \end{array} \right)  \nonumber \\ 
R_{w_5}(v) &=  \left( \begin{array}{ccc}  0  &  0  &  1  \\  0  &  -1  &  0  \\  1  &  0  &  0  \\  \end{array} \right) v + \left( \begin{array}{c} 1/4 \\ 1/4 \\ -1/4 \end{array} \right)  \ \ \ \
& R_{w_6}(v) &=  \left( \begin{array}{ccc}  0  &  1  &  0  \\  1  &  0  &  0  \\  0  &  0  &  -1  \\  \end{array} \right) v + \left( \begin{array}{c} -1/4 \\ 1/4 \\ 1/4 \end{array} \right)  \,.
\end{align}

The invariant momenta are lines in momentum space parametrized by a paramter $\alpha$ and are given by:
\begin{align}
w_1 &: (0,\alpha,-\alpha) &w_2 &: (\alpha,0,-\alpha) &w_3 &: (\alpha,-\alpha,0) \nonumber \\
w_4 &: (0,\alpha,\alpha) &w_5 &: (\alpha,0,\alpha) &w_6 &: (\alpha,\alpha,0)
\end{align}

The complete list of how the symmetries act on the operators in the unit cell can then be summarized as follows:
\begin{equation}
\begin{array}{lll} \hline & R_{w_1} & \\ \hline\gamma_{0} \rightarrow \gamma_{6}, o_{0} = (0,0,-1)^T  & \gamma_{1} \rightarrow \gamma_{5}, o_{1} = (0,0,0)^T  & \gamma_{2} \rightarrow \gamma_{4}, o_{2} = (0,0,0)^T  \\ \gamma_{3} \rightarrow \gamma_{7}, o_{3} = (0,0,0)^T  & \gamma_{4} \rightarrow \gamma_{2}, o_{4} = (0,0,-1)^T  & \gamma_{5} \rightarrow \gamma_{1}, o_{5} = (0,0,0)^T  \\ \gamma_{6} \rightarrow \gamma_{0}, o_{6} = (0,0,0)^T  & \gamma_{7} \rightarrow \gamma_{3}, o_{7} = (0,0,0)^T  & \gamma_{8} \rightarrow \gamma_{10}, o_{8} = (0,0,0)^T  \\ \gamma_{9} \rightarrow \gamma_{9}, o_{9} = (0,0,0)^T  & \gamma_{10} \rightarrow \gamma_{8}, o_{10} = (0,0,0)^T  & \gamma_{11} \rightarrow \gamma_{11}, o_{11} = (0,0,-1)^T  \\ \hline & R_{w_2} & \\ \hline\gamma_{0} \rightarrow \gamma_{2}, o_{0} = (0,0,0)^T  & \gamma_{1} \rightarrow \gamma_{1}, o_{1} = (0,0,0)^T  & \gamma_{2} \rightarrow \gamma_{0}, o_{2} = (0,0,0)^T  \\ \gamma_{3} \rightarrow \gamma_{3}, o_{3} = (0,0,0)^T  & \gamma_{4} \rightarrow \gamma_{10}, o_{4} = (0,0,0)^T  & \gamma_{5} \rightarrow \gamma_{9}, o_{5} = (0,0,0)^T  \\ \gamma_{6} \rightarrow \gamma_{8}, o_{6} = (0,1,0)^T  & \gamma_{7} \rightarrow \gamma_{11}, o_{7} = (0,0,-1)^T  & \gamma_{8} \rightarrow \gamma_{6}, o_{8} = (0,0,0)^T  \\ \gamma_{9} \rightarrow \gamma_{5}, o_{9} = (0,0,0)^T  & \gamma_{10} \rightarrow \gamma_{4}, o_{10} = (0,0,0)^T  & \gamma_{11} \rightarrow \gamma_{7}, o_{11} = (0,0,0)^T  \\ \hline & R_{w_3} & \\ \hline\gamma_{0} \rightarrow \gamma_{10}, o_{0} = (0,0,0)^T  & \gamma_{1} \rightarrow \gamma_{9}, o_{1} = (0,0,-1)^T  & \gamma_{2} \rightarrow \gamma_{8}, o_{2} = (0,0,-1)^T  \\ \gamma_{3} \rightarrow \gamma_{11}, o_{3} = (0,0,-1)^T  & \gamma_{4} \rightarrow \gamma_{6}, o_{4} = (0,0,-1)^T  & \gamma_{5} \rightarrow \gamma_{5}, o_{5} = (0,0,0)^T  \\ \gamma_{6} \rightarrow \gamma_{4}, o_{6} = (0,0,-1)^T  & \gamma_{7} \rightarrow \gamma_{7}, o_{7} = (0,0,-1)^T  & \gamma_{8} \rightarrow \gamma_{2}, o_{8} = (0,0,-1)^T  \\ \gamma_{9} \rightarrow \gamma_{1}, o_{9} = (0,0,-1)^T  & \gamma_{10} \rightarrow \gamma_{0}, o_{10} = (0,0,0)^T  & \gamma_{11} \rightarrow \gamma_{3}, o_{11} = (0,0,-1)^T  \\ \hline & R_{w_4} & \\ \hline\gamma_{0} \rightarrow \gamma_{7}, o_{0} = (0,0,0)^T  & \gamma_{1} \rightarrow \gamma_{4}, o_{1} = (0,0,0)^T  & \gamma_{2} \rightarrow \gamma_{5}, o_{2} = (0,0,0)^T  \\ \gamma_{3} \rightarrow \gamma_{6}, o_{3} = (0,0,-1)^T  & \gamma_{4} \rightarrow \gamma_{1}, o_{4} = (0,0,0)^T  & \gamma_{5} \rightarrow \gamma_{2}, o_{5} = (0,0,-1)^T  \\ \gamma_{6} \rightarrow \gamma_{3}, o_{6} = (0,0,0)^T  & \gamma_{7} \rightarrow \gamma_{0}, o_{7} = (0,0,0)^T  & \gamma_{8} \rightarrow \gamma_{8}, o_{8} = (0,0,0)^T  \\ \gamma_{9} \rightarrow \gamma_{11}, o_{9} = (0,0,-1)^T  & \gamma_{10} \rightarrow \gamma_{10}, o_{10} = (0,0,0)^T  & \gamma_{11} \rightarrow \gamma_{9}, o_{11} = (0,0,-1)^T  \\ \hline & R_{w_5} & \\ \hline\gamma_{0} \rightarrow \gamma_{0}, o_{0} = (0,0,0)^T  & \gamma_{1} \rightarrow \gamma_{3}, o_{1} = (0,0,0)^T  & \gamma_{2} \rightarrow \gamma_{2}, o_{2} = (0,-1,0)^T  \\ \gamma_{3} \rightarrow \gamma_{1}, o_{3} = (0,0,0)^T  & \gamma_{4} \rightarrow \gamma_{11}, o_{4} = (0,0,-1)^T  & \gamma_{5} \rightarrow \gamma_{8}, o_{5} = (0,0,0)^T  \\ \gamma_{6} \rightarrow \gamma_{9}, o_{6} = (0,0,0)^T  & \gamma_{7} \rightarrow \gamma_{10}, o_{7} = (0,-1,0)^T  & \gamma_{8} \rightarrow \gamma_{5}, o_{8} = (0,0,0)^T  \\ \gamma_{9} \rightarrow \gamma_{6}, o_{9} = (0,0,-1)^T  & \gamma_{10} \rightarrow \gamma_{7}, o_{10} = (0,-1,0)^T  & \gamma_{11} \rightarrow \gamma_{4}, o_{11} = (0,0,0)^T  \\ \hline & R_{w_6} & \\ \hline\gamma_{0} \rightarrow \gamma_{9}, o_{0} = (0,0,0)^T  & \gamma_{1} \rightarrow \gamma_{10}, o_{1} = (0,0,0)^T  & \gamma_{2} \rightarrow \gamma_{11}, o_{2} = (0,0,-1)^T  \\ \gamma_{3} \rightarrow \gamma_{8}, o_{3} = (0,0,0)^T  & \gamma_{4} \rightarrow \gamma_{4}, o_{4} = (0,0,0)^T  & \gamma_{5} \rightarrow \gamma_{7}, o_{5} = (0,0,0)^T  \\ \gamma_{6} \rightarrow \gamma_{6}, o_{6} = (0,0,-1)^T  & \gamma_{7} \rightarrow \gamma_{5}, o_{7} = (0,0,0)^T  & \gamma_{8} \rightarrow \gamma_{3}, o_{8} = (0,0,0)^T  \\ \gamma_{9} \rightarrow \gamma_{0}, o_{9} = (0,0,0)^T  & \gamma_{10} \rightarrow \gamma_{1}, o_{10} = (0,0,0)^T  & \gamma_{11} \rightarrow \gamma_{2}, o_{11} = (0,0,-1)^T  \\  \end{array} 
\end{equation}
The operators $U^\alpha(k)$, for $\alpha=w_1, \ldots, w_6$, can be constructed from the table above. Inspection of the resulting matrices shows that in
each case, they take the form $U = e^{i \phi} P$ with $P$ a permutation matrix with signature 2. Therefore, at the momenta invariant under each symmetry,
gapless excitations are guaranteed.

\end{widetext}

%%%%%%%%%%%%%%%%%%%%%%%%%%%%%%%%%%%%%%%%%%%%%%%%%%%%%%%%%

\section{Matrix product states for projected wave functions}
\label{app:numerics}

Our numerical analysis of the Gutzwiller-projected BCS wavefunctions is performed within a matrix-product state (MPS) formalism. While
these results are in principle also accessible to sign-free variational Monte Carlo calculations, the MPS approach allows easy computation
of bipartite entanglement entropies and thus connects most easily to the results presented in Sec.~\ref{sec:quasi1dw2}.

In a first step, we obtain an MPS representation of the desired free-fermion wavefunction, i.e. the ground state of Eqn.~\eqnref{eqn:ManyMaj},
using a recently developed technique described in detail in Ref.~\onlinecite{fishman2015}. This approach starts from the correlation matrix
$C_{ij} = \langle c_i^\dagger c_j \rangle$, and iteratively diagonalizes this matrix through series of local orthogonal transformations. The MPS
representation is then built by applying these same orthogonal transformations (in reverse order) to a simple initial product state. Since these
are local unitary transformations, every step can be approximately performed using standard time-evolution techniques akin to the
TEBD method~\cite{vidal2003-1,daley2004,feiguin2004}.

In a next step, we apply a projection operator that selects only the states with exactly one fermion per site. Since this is a local projection,
it can be done to each tensor in the MPS independently by multiplying the physical index with the appropriate projector, Eqn.~\eqnref{eqn:gutzproj}.
This step can be done without further approximation. Throughout this procedure, care must be taken to correctly implement the fermionic
anticommutation rules in order to obtain a spin wavefunction with the correct sign structure. Finally, after having performed the projection,
expectation values and bipartite entanglement entropies can be computed using standard MPS techniques.

%%%%%%%%%%%%%%%%%%%%%%%%%%%%%%%%%%%%%%%%%%%%%%%%%%%%%%%%%

\bibliography{spg}

\end{document}